\documentclass[11pt, fleqn, a4paper]{article}

\usepackage[utf8]{inputenc}
\usepackage[american]{babel}
\usepackage{amsmath, amssymb, amsthm}

\usepackage[backend=biber, style=authoryear, uniquename=false, uniquelist=false]{biblatex}
\usepackage{geometry, setspace}
\AtBeginEnvironment{tabular}{\doublespacing} % table spacing
\AtEndEnvironment{tabular}{\vspace{6pt}}
\AtBeginEnvironment{longtable}{\doublespacing} % longtable spacing
\AtEndEnvironment{longtable}{\vspace{6pt}}

\usepackage{appendix}

\usepackage{graphicx, subfigure} % figures
\usepackage{booktabs, longtable} % tables
\usepackage{caption, placeins} % captions and placement of figures and tables
\usepackage{hyperref} % hyperlinks
\hypersetup{
	colorlinks=true,
	urlcolor=blue,
	linkcolor=red,
	citecolor=blue
}
\usepackage[nameinlink]{cleveref} % table, figure, and section references
\newcounter{cTable}
\crefname{cTable}{table}{tables}
\Crefname{CTable}{Table}{Tables}
\crefname{equation}{equation}{equations}
\Crefname{equation}{Equation}{Equations}

\usepackage{orcidlink} % orcid symbol
\usepackage{authblk} % affiliations

\begin{document}	
	
	\title{Effect of Cigarette Price and Tax Increases on Smoking in Europe: A Difference-in-Differences Study with Double Machine Learning\thanks{Corresponding author: Andreas Stoller, University of Fribourg, Department of Economics, Bd de Perolles 90, 1700 Fribourg, Switzerland, andreas.stoller@unifr.ch}}
	
	\date{April 7, 2026 \vspace{-1cm}}
	
	\author[]{Andreas Stoller$^{1}$ \orcidlink{0000-0002-2404-648X} \& Martin Huber$^{1}$ \orcidlink{0000-0002-8590-9402}}
	\affil[]{$^{1}$University of Fribourg, Department of Economics \vspace{-0.5cm}}
	
	\maketitle
	
	\begin{abstract}
	
		We estimate the effect of cigarette price and tax increases on smoking rates using Eurobarometer survey data from 27 European Union countries between 2012 and 2020. Following a difference-in-differences approach, we compare individuals exposed to large price and tax increases with those in stable price and tax environments. Estimation is based on a difference-in-differences estimator with double machine learning, which relaxes the functional form assumptions typically imposed by parametric approaches such as two-way fixed effects. Our results indicate that tax increases reduce smoking rates among individuals who smoke at least once per month and among daily smokers. The reduction is primarily driven by individuals aged 15--24. We examine the sensitivity of our findings to functional form assumptions and treatment definitions. While estimates are robust to alternative functional form assumptions, they are sensitive to whether the treatment is defined as binary or continuous.
		
		\vspace{0.5cm}
		
		
		\vspace{0.5cm}
		
		
	\end{abstract}

\newpage

\section{Introduction}

The tobacco epidemic remains a major public health challenge despite great prevention efforts. Although global smoking rates have declined substantially over recent decades \parencite{Reitsma21}, smoking still accounts for 8.71 million deaths worldwide, which is about 15.4\% of all deaths \parencite{Murray20}. Europe has the second-highest smoking rate globally \parencite{WHO24}, and roughly 850,000 deaths each year in the region are attributable to smoking \parencite{EU24}. Smoking therefore remains one of the leading preventable risk factors, particularly in Europe.

Tobacco taxation is a central policy tool for reducing tobacco use and is widely applied in Europe and globally. By signing the “WHO Framework Convention on Tobacco Control,” 168 countries support price and tax policies as effective instruments to reduce tobacco consumption \parencite{WHO03}. Moreover, European Union (EU) member states are required to levy tobacco taxes amounting to at least 60\% of the retail selling price of cigarettes \parencite{EU11}.

At the same time, new tobacco and nicotine products have emerged, most notably electronic cigarettes (e-cigarettes). The share of e-cigarette users in the general population of Organisation for Economic Co-operation and Development (OECD) member countries increased from 3\% in 2016 to 6\% in 2023 \parencite{OECD25}. Among adolescents, use of these products is even more widespread. Around 20\% of 15-year-olds reported e-cigarette use in the past month in 2022. Evidence from the Health Behaviour in School-aged Children survey, covering 44 mostly European countries, further shows that more than 18\% of adolescents have tried e-cigarettes at least once, and about 10\% reported use in the past 30 days around the same time period \parencite{Charrier24}. The introduction and rising uptake of these products add a new layer of complexity when evaluating policies targeting traditional tobacco products such as cigarettes.

Tobacco taxes generally raise tobacco prices, and demand for traditional cigarettes responds to these price changes \parencite{DeCicca22}. In their extensive literature review, which incorporates recent quasi-experimental evidence, \textcite{DeCicca22} conclude that cigarette demand is relatively inelastic, with overall price elasticities of around -0.5 across both the extensive and intensive margins. Some studies also examine whether the effectiveness of tobacco taxation declines over time, as tax and price levels rise, for example because remaining smokers may become less price responsive or tax avoidance may increase. Indeed, several studies suggest that the impact of tobacco taxes on smoking behavior diminishes, or may even disappear, over time \parencite{Bishop18, Hansen17, Shrestha25}. However, existing evidence remains limited in some respects, as it is predominantly based on the United States and focuses on periods preceding the widespread uptake of e-cigarettes.

The evolution of tobacco taxation, together with the emergence of new nicotine and tobacco products, warrants ongoing evaluation of tax policy. On the one hand, tobacco taxation may lose effectiveness once tax and price levels become sufficiently high. On the other hand, the availability of new nicotine and tobacco products may result in unintended effects if higher cigarette taxes lead to substitution toward these products. Studying tobacco taxation in this context therefore remains crucial.

On a more methodological note, current estimates are most often based on conventional difference-in-differences (DiD) estimators. Typically, a two-way fixed effects (TWFE) specification regresses the smoking outcome on continuous tobacco prices and taxes, along with confounding factors and additional fixed effects. These estimates rely on strong functional form assumptions. Most importantly, they assume that the treatment effect is homogeneous across all values of the treatment, even though effects at high tax or price levels may differ from those at lower levels, and the impact of a decrease in prices or taxes may not mirror the effect of an increase. In addition, these models assume that confounding factors influence the outcome in a linear and additive way, which may be unrealistic. For example, an additional year of age may affect the smoking behavior of a young individual differently than that of an older individual.

Recent methodological advancements offer new opportunities to improve our understanding of how tobacco prices and taxes affect smoking. Double machine learning (DML) estimators incorporate machine learning methods such as lasso and random forests into causal analysis \parencite{Chernozhukov18}, allowing for more flexible control of confounding factors than commonly used parametric approaches. These estimators are also doubly robust and \textcite{Neyman59} orthogonal, allowing limited approximation errors in treatment and outcome models under certain regularity conditions. Building on the DML framework, \textcite{Zimmert20} and \textcite{Chang20} propose novel DiD estimators that integrate DML. The DiD estimator with DML (DiDDML) by \textcite{Zimmert20} is particularly suitable for repeated cross-sections, where sampled individuals may change over time.

The methodological choices in the current literature on tobacco prices and taxes may influence reported results. This study focuses on two methodological aspects. First, the treatment effect may not be homogeneous across all price and tax values. Instead of treating prices and taxes as continuous variables, we use binary indicators and compare treatment groups experiencing increases in prices and taxes with control groups in which prices and taxes remain stable. Second, confounding factors may influence smoking outcomes in ways that are neither additive nor linear. To allow for more flexible, nonparametric relationships, we apply a DiDDML estimator rather than conventional parametric DiD estimators. 

Our study estimates the effect of cigarette price and tax increases on smoking rates using survey data from 27 European countries between 2012 and 2020. The main estimates rely on the DiDDML estimator by \textcite{Zimmert20}. Our findings show that a cigarette tax increase reduces the smoking rate by 3.44 percentage points (pp) (a 15\% reduction; p $=0.04$) in the post-treatment period among individuals who experience a tax increase, considering smokers who smoke at least once per month. For daily smokers, the smoking rate falls by 3.15 pp (a 15\% reduction; p $=0.09$). However, we do not find a statistically significant effect of cigarette price increases at the 5\% level, which may be due to upward bias because of endogeneity in price changes. 

The contribution of this study is both substantive and methodological. First, we provide new evidence on the effects of tobacco prices and taxes using recent data from European countries operating in high-tax environments. Most importantly, tobacco taxation is examined under these new circumstances. Moreover, we contribute to the limited body of quasi-experimental evidence from Europe. Second, we compare estimates from flexible causal machine learning methods with those obtained from conventional parametric DiD approaches, allowing us to assess the sensitivity of results to functional form assumptions. In addition, we explore and discuss the implications of using binary versus continuous treatment definitions.

The paper is organized as follows. \Cref{sec:literature} reviews the relevant literature. \Cref{sec:data} presents the survey and policy data, while \cref{sec:method} introduces the identification and empirical strategy, as well as the estimation procedure. The results are presented in \cref{sec:results}, followed by robustness checks in \cref{sec:robustChecks}. \Cref{sec:conclusion} concludes.

\section{Literature}
\label{sec:literature}

\textcite{Chaloupka00} as well as \textcite{Gallet03} review studies conducted before 2000, concluding that the price elasticity of cigarette demand is negative and inelastic. These studies also find that the extensive margin (smoking participation) and the intensive margin (cigarettes smoked conditional on participation) contribute equally to overall price elasticity. However, this early literature does not yet incorporate methodological advances from modern quasi-experimental designs. 

\textcite{DeCicca22} review more recent work relying on quasi-experimental approaches and report similar conclusions, with extensive margin elasticities typically ranging from -0.1 to -0.3. This implies that a 10\% increase in cigarette prices reduces the smoking rate by 1--3\%. Evidence from studies conducted both before and after the 2000s suggests that overall price elasticities are approximately -0.5, considering both the extensive and intensive margin.

The most recent quasi-experimental studies use price as well as tax variation across time and regions to estimate the price and tax elasticity of cigarette demand. These studies usually rely on survey or purchase data for individuals or households. Most find negative elasticities, indicating that higher cigarette taxes reduce the smoking rate and the number of cigarettes consumed \parencite{Bishop18, Cotti16, Cotti18, Maclean14, Nesson17a, Nesson17b, Pesko16, Pesko20, Shrestha25}. However, some studies suggest that the impact of tobacco taxes on smoking weakens or even disappears over time \parencite{Bishop18, Hansen17, Shrestha25}. One study by \textcite{Callison14} even finds elasticities close to zero or null effects.

Multiple studies find that light smokers are more responsive to price increases than heavy smokers \parencite{Maclean14, Nesson17b, Cotti18}. These findings support the hypothesis that the gradual crowding out of more price-responsive smokers contributes, at least in part, to the diminishing effectiveness of tobacco taxation over time. Another relevant factor may be increased tax avoidance through illicit or cross-border cigarette purchases, as discussed in \textcite{Bishop18}. Several studies also show that higher traditional cigarette taxes reduce cigarette use while increasing e-cigarette use among adults, suggesting that cigarettes and e-cigarettes are substitutes \parencite{Cotti18, Pesko20}, although \textcite{Cotti16} provide evidence that they may be partially economic complements. With the emergence of new nicotine and tobacco products, substitution may therefore become more prevalent. Consequently, cigarette taxes may become less effective due to saturation effects, yet more effective through unintended substitution and tax avoidance. Hence, the expected change in the effectiveness of tobacco taxes on cigarette consumption over time is not straightforward. 

The evidence on effect heterogeneity of price and tax increases is mixed. Effect heterogeneity by gender, age, education, and income is commonly examined. \textcite{Maclean14} find that women and individuals with higher levels of education are most responsive, while \textcite{Pesko16} report that young individuals and those with higher income and education drive the effects. In contrast, \textcite{Pesko20} suggest that the effects are driven by women. However, several studies find no notable differences across one or multiple subgroups \parencite{Callison14, Cotti16, Cotti18, Pesko20}.

European evidence on the impact of price and tax increases is scarce, as most existing research is based on data from the United States. To the best of our knowledge, the only quasi-experimental study examining their effects on smoking in Europe is the working paper by \textcite{Odermatt18}. Using an instrumental variables approach with taxes as instruments for cigarette prices, they find negative but insignificant effects on the smoking rate and the number of cigarettes consumed across 21 European countries from 1990 to 2012. The effects are driven by men and individuals younger than 30.

Regarding estimation, the literature on price and tax elasticities of tobacco demand relies mostly on TWFE estimators, for example \textcite{Cotti16, Cotti18, Pesko20}. Hence, these estimators are parametric. The outcome variable is either binary, indicating whether an individual smokes, or continuous, such as the number of cigarettes consumed. The treatments are typically cigarette prices or taxes, expressed in monetary units and included as continuous regressors. Causal effects are identified by exploiting geographic and temporal variation in these price and tax measures. In addition, these models commonly include time and region fixed effects, along with sets of demographic and policy covariates, to account for confounding factors. These analyses are usually conducted using large individual- or household-level datasets spanning numerous quarters or years.

Cigarette prices and taxes are typically measured as continuous units, and elasticities of demand are conveniently estimated by including them as continuous treatments in TWFE models. Such approaches implicitly assume homogeneous treatment effects of price and tax changes. However, the effectiveness of small versus large increases, or of increases at low versus high price or tax levels, may differ. One of the few exceptions is \textcite{Callison14}, who explicitly investigate dose responses by focusing on the effects of large tax increases.

Identification of causal effects with continuous treatments is not straightforward. In the conventional DiD design, the treatment is binary, with only the treated group exposed to the intervention while the control group is not. In contrast, prices and taxes vary continuously over time and across regions, making a clean comparison between treated and untreated groups infeasible. As a result, it is not possible to test the parallel trends assumption in conventional ways when treatments remain continuous. Some studies construct binary treatments to approximate a more conventional DiD framework and to verify parallel trends to some extent, as in \textcite{Pesko20}.

Recent advancements in the DiD literature also call for a reassessment of existing studies on cigarette price and tax increases. TWFE estimators may be biased in settings with multiple time periods and staggered treatment timing \textcite{GoodmanBacon21}. Their estimates can be biased when treatment effects vary across treatment groups and time. Because of this so-called “negative weights” bias, several estimators have been developed that are robust to treatment-effect heterogeneity \parencite{Callaway21, DeChaisemartin20, Sun21, Borusyak24}. \textcite{Shrestha25} examines this issue within the cigarette demand elasticity literature by estimating the effect of cigarette taxes using both TWFE and heterogeneity-robust staggered DiD estimators. The results indicate that TWFE estimates are biased toward zero, implying that previous elasticity estimates based on TWFE may underestimate the effectiveness of cigarette taxes in reducing smoking.

This study contributes to the literature in several ways. First, we deviate from common modeling choices in multiple respects. Rather than using continuous tax and price variables, we compare treatment groups that experience a price or tax increase with control groups that remain in stable price and tax environments. The treatment is therefore binary and more consistent with the conventional DiD framework. Second, instead of relying on parametric models, we apply causal machine learning approaches that allow for nonparametric confounding. Finally, we compare estimates based on these new modeling choices with those obtained from more conventional parametric specifications. While we do not investigate the “negative weights” problem directly, it is important to note that our study design is robust to this problem, as discussed in \cref{sec:results}. Beyond the methodological contribution, we provide empirical evidence for the European population in high-tax environments and in the presence of emerging products such as e-cigarettes. We assess the effectiveness of price and tax increases under these new circumstances, thereby updating the evidence on these regulatory levers for current tobacco policy.

\section{Data}
\label{sec:data}

We use European survey data combined with information on tobacco prevention policies, cigarette prices, and cigarette taxes. The repeated cross-section survey data from the Eurobarometer provide smoking status and socio-demographic characteristics of individuals \parencite{EB14, EB18, EB19, EB21}. We supplement these data with country-level cigarette price and tax information from \textcite{WHO21taxes, WHO22prices}. Additionally, the MPOWER indicators measure the intensity of various tobacco prevention policies at the country level \parencite{WHO23mpowerP, WHO23mpowerO, WHO23mpowerW, WHO23mpowerE}. The final dataset contains 107,258 individuals from 27 EU member countries in 2012, 2014, 2018, and 2020. Observations from Croatia are excluded because the country is not present in all survey waves. We also exclude 0.2\% of observations due to unknown smoking status.

\subsection{Treatment: Cigarette Price and Tax-Share Increases}
\label{sec:dataTreatment}

\begin{figure}[htbp!]
	\centering
	\subfigure[Distribution of prices, in PPP\$]{\includegraphics[width=0.45\textwidth]{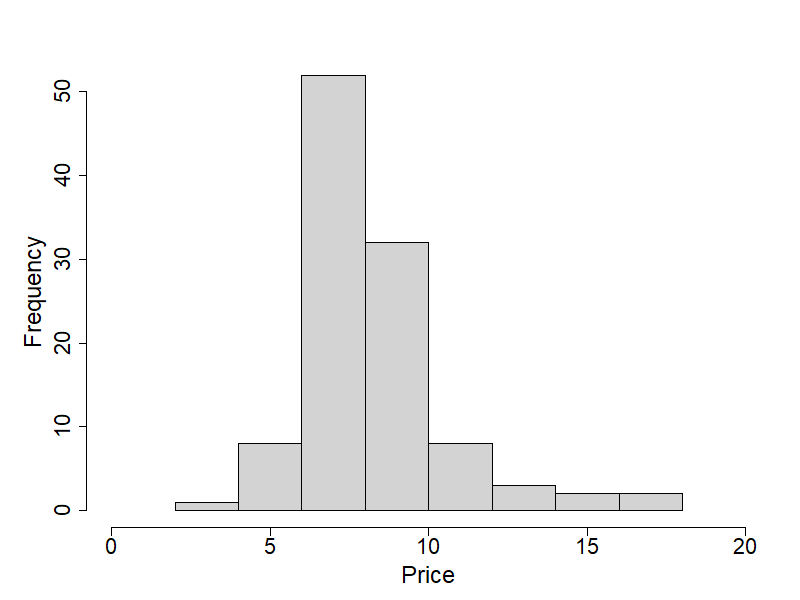}}
	\subfigure[Distribution of tax, as share of price]{\includegraphics[width=0.45\textwidth]{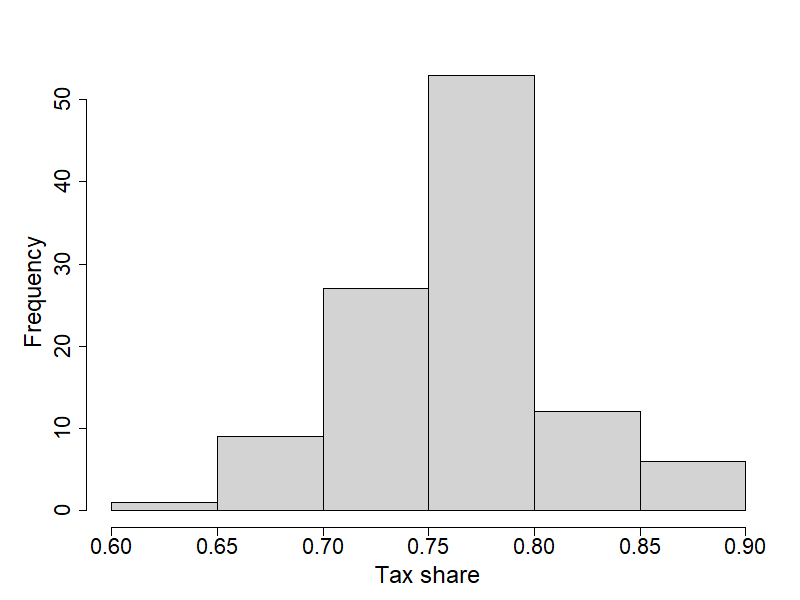}}
	\caption{Descriptive information on cigarette prices and tax shares}
	\label{fig:cigLevels}
	\caption*{\footnotesize \textit{Note: Figure (a) shows the distribution of cigarette prices and (b) displays the distribution of cigarette tax shares, considering the full sample.}}
\end{figure}

The cigarette price data refer to the retail price of a pack of 20 cigarettes of the most popular brand \parencite{WHO22prices}, adjusted for purchasing power parity (PPP) and inflation. The cigarette tax data capture the total tax share of a pack of 20 cigarettes of the most popular brand \parencite{WHO21taxes}. The total tax share is defined as the tax as a percentage of the retail price, including amount-specific excise taxes, ad valorem excise taxes, import duties, and value added tax. Both price and tax-share measures are collected at the country level. \Cref{fig:cigLevels} shows the distribution of cigarette prices and tax shares across all available countries. Most cigarette prices fall between 5--10 PPP\$, while the tax share most often ranges between 75--80\%.

\begin{figure}[htbp!]
	\centering
	\subfigure[Distribution of changes in prices]{\includegraphics[width=0.45\textwidth]{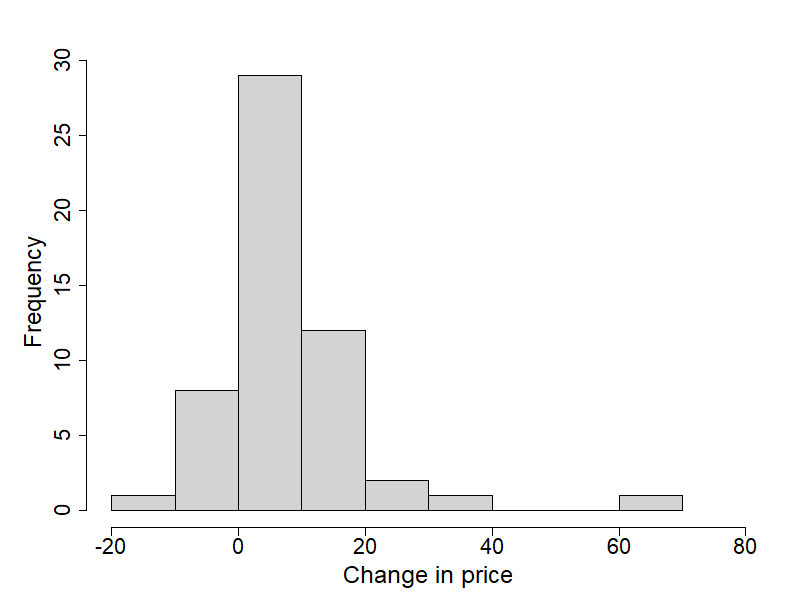}}
	\subfigure[Distribution of changes in tax shares]{\includegraphics[width=0.45\textwidth]{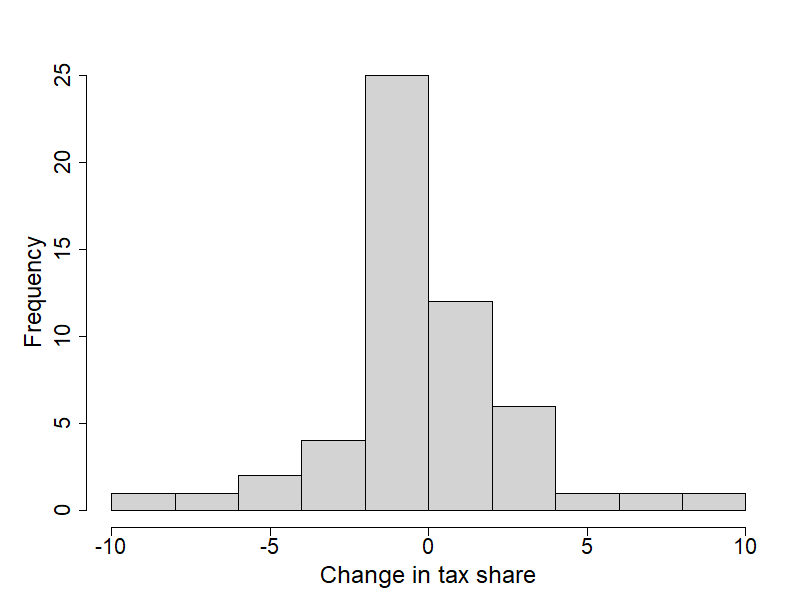}}
	\caption{Descriptive information on changes in cigarette prices and tax shares, in percent}
	\label{fig:cigChanges}
	\caption*{\footnotesize \textit{Note: Figure (a) shows the distribution of cigarette price changes between the pre- and post-treatment periods and (b) displays the distribution of cigarette tax share changes between the pre- and post-treatment periods.}}
\end{figure}

We are going to assess the effectiveness of price and tax-share increases over two periods, 2012--2014 and 2018--2020. For the first period, 2012 serves as the pre-treatment and 2014 as the post-treatment period. Similarly, for the second period, 2018 serves as the pre-treatment and 2020 as the post-treatment period. To evaluate increases in cigarette prices and in the tax shares, we create binary treatment variables. The treatment group includes individuals from countries experiencing substantial price or tax-share increases, while the control group consists of individuals from countries with stable prices or tax shares. 

For prices, the treatment group includes individuals from countries where the cigarette price increases by more than 15\% from the pre- to post-treatment period, while the control group consists of individuals from countries with stable prices (a change of $\pm$5\%). Regarding taxes, the treatment group includes individuals from countries where the tax-share increases by more than 2\%, and the control group consists of individuals from countries where the tax share remains unchanged. As a result, individuals from countries not assigned to either the control or treatment group are excluded from the analysis. 

\Cref{fig:cigChanges} presents changes in cigarette prices and tax shares between the pre- and post-treatment periods. \Cref{tab:treatmentsPrices} and \cref{tab:treatmentsTaxes} summarize cigarette prices and tax shares, changes in these prices and taxes, as well as the corresponding treatment assignments, by country. 

\Cref{tab:changePricesTaxes} presents the average change in prices and tax shares from the pre- to the post-treatment period within the control and treatment groups. For prices, the treatment groups experience an average increase of 27.20\%, whereas the control groups show almost no change. For taxes, the treatment groups exhibit an average increase in the tax share of 4.13\%, while the control groups show no change in the tax share.

\begin{table}[htbp!]
	\caption{Changes in prices and tax shares, in percent}
	\label{tab:changePricesTaxes}
	\centering
	\begin{tabular}{p{.15\textwidth}p{.15\textwidth}p{.15\textwidth}p{.15\textwidth}p{.15\textwidth}}
		& \multicolumn{2}{c}{Price increase} & \multicolumn{2}{c}{Tax-share increase} \\
		\cmidrule(lr){2-3} \cmidrule(lr){4-5}
		Period & Control & Treated & Control & Treated \\
		\hline 
		2012--2014 & 0.86 & 28.87 & 0.00 & 2.53 \\ 
		2018--2020 & 0.20 & 24.43 & 0.00 & 5.41 \\ 
		Overall & 0.37 & 27.20 & 0.00 & 4.13 \\ 
		\hline\hline
	\end{tabular}
	\caption*{\footnotesize \textit{Note: This table reports average changes in prices and tax shares in percent from the pre- to post-treatment period for control and treatment groups, separately for the periods 2012--2014, 2018--2020, and across all periods.}}
\end{table}

\FloatBarrier

\subsection{Outcome: Smoking Rate}

The outcome variable is the individual's smoking status, defined as a binary indicator equal to one if the individual smokes cigarettes and zero otherwise. Consequently, we estimate the effect of increases in cigarette prices and in the tax share on the extensive margin, i.e., the smoking rate. We analyze effects separately for individuals who smoke at least once per month and for daily smokers. We thus have two separate outcomes, the smoking rate of current smokers and the smoking rate of daily smokers. The smoking rate of current smokers therefore includes both daily and occasional smokers. We do not consider the smoking rate of occasional smokers as a separate outcome because relatively few individuals fall into this category, which could lead to floor effects given that their smoking rate is close to zero. Data on the number of cigarettes consumed are not available for all periods, preventing an analysis of the intensive margin.

\Cref{tab:smokingRateInTreatmentGroups} shows the development of smoking rates within the control and treatment groups from the pre- to the post-treatment periods. The table presents the four main analyses, reporting current and daily smoking rates for both the increases in prices and tax shares. For price increases, individuals exposed to a price increase exhibit substantially higher smoking rates than the control group in the pre-treatment period, and the subsequent decrease in both smoking rates is more pronounced in the treatment group. For tax-share increases, pre-treatment smoking rates are at similar levels across the control and treatment groups, yet the decline in both current and daily smoking rates is again more pronounced among individuals experiencing an increase in the tax share.

\begin{table}[htbp!]
	\caption{Smoking rates across treatment groups and time periods}
	\label{tab:smokingRateInTreatmentGroups}
	\centering
	\begin{tabular}{p{0.1\textwidth}p{0.1\textwidth}p{0.1\textwidth}p{0.09\textwidth}p{0.09\textwidth}p{0.09\textwidth}p{0.09\textwidth}p{0.09\textwidth}p{0.09\textwidth}}
		& & & \multicolumn{2}{c}{Pre-treatment} & \multicolumn{2}{c}{Post-treatment} & \\ 
		\cmidrule(lr){4-5} \cmidrule(lr){6-7}
		Treatment & Outcome & Status & Rate & N & Rate & N & Difference \\ 
		\hline
		Cigarette & Current & D $=$ 0 & 0.27 & 11,831 & 0.24 & 25,757 & -0.03 \\ 
		 prices  & & D $=$ 1 & 0.31 & 10,142 & 0.18 & 6,161 & -0.13 \\ 
		  & Daily & D $=$ 0 & 0.26 & 11,831 & 0.22 & 25,757 & -0.04 \\ 
		  & & D $=$ 1 & 0.30 & 10,142 & 0.16 & 6,161 & -0.14 \\ 
		Cigarette & Current & D $=$ 0 & 0.25 & 15,318 & 0.22 & 21,569 & -0.03 \\ 
		 taxes  &   & D $=$ 1 & 0.26 & 8,072 & 0.20 & 10,370 & -0.06 \\ 
		  & Daily & D $=$ 0 & 0.24 & 15,318 & 0.21 & 21,569 & -0.03 \\ 
		  &   & D $=$ 1 & 0.25 & 8,072 & 0.18 & 10,370 & -0.07 \\
		\hline\hline
	\end{tabular}
	\caption*{\footnotesize \textit{Note: This table presents the smoking rates and the number of observations, grouped by control and treatment groups and by pre- and post-treatment periods. The last column reports the change in smoking rates before and after the price and tax-share increases.}}
\end{table}

\Cref{tab:smokingHetGroups} reports smoking rates within the samples defined according to the definitions of price and tax treatments above, by socio-demographic groups. Men have substantially higher smoking rates than women across all specifications. Smoking rates are highest among individuals aged 25--44 relative to other age groups. Individuals who completed their last education between ages 16--19, a proxy for a medium level of education, display the highest smoking rates compared to those with lower or higher education levels.

\begin{table}[htbp!]
	\caption{Smoking rates, by socio-demographic groups}
	\label{tab:smokingHetGroups}
	\centering
	\begin{tabular}{p{.3\textwidth}p{.1\textwidth}p{.1\textwidth}p{.1\textwidth}p{.1\textwidth}}
		& \multicolumn{2}{c}{Price increase} & \multicolumn{2}{c}{Tax-share increase} \\
		\cmidrule(lr){2-3} \cmidrule(lr){4-5}
		Group & Current & Daily & Current & Daily \\
		\hline 
		Men & 0.29 & 0.28 & 0.27 & 0.26 \\ 
		Women & 0.22 & 0.20 & 0.20 & 0.18 \\ 
		15--24 years old & 0.26 & 0.24 & 0.24 & 0.22 \\ 
		25--44 years old & 0.32 & 0.31 & 0.30 & 0.29 \\ 
		45--64 years old & 0.29 & 0.28 & 0.26 & 0.25 \\ 
		65+ years old & 0.12 & 0.11 & 0.11 & 0.10 \\ 
		Age last education $\leq$15 & 0.23 & 0.23 & 0.22 & 0.21 \\ 
		Age last education 16--19 & 0.32 & 0.30 & 0.29 & 0.28 \\ 
		Age last education 20+ & 0.20 & 0.18 & 0.18 & 0.17 \\ 
		\hline\hline
	\end{tabular}
	\caption*{\footnotesize \textit{Note: This table shows smoking rates for the control and treatment groups across the samples defined by the different price and tax-share increase treatments, by socio-demographic groups. This includes men and women, the age groups 15--24, 25--44, 45--64, and 65+, as well as individuals who completed full-time education at ages up to 15, 16--19, and 20+.}}
\end{table}

\FloatBarrier

\subsection{Covariates: Socio-Demographics and Tobacco Prevention Policies}

We control for treatment history to ensure that individuals are compared within similar price and tax environments, using pre-treatment cigarette prices in the price analysis and pre-treatment tax shares in the tax-share analysis as control variables. We also include individuals’ socio-demographic characteristics as control variables. These characteristics include age, gender, household composition (i.e., number of persons living in the household), foreign nationality (i.e., a dummy indicating foreign nationality), age at completion of full-time education, marital status, type of community (i.e., degree of urbanization), and occupation.

Additionally, the MPOWER indicators cover six dimensions of tobacco prevention policy at the country level to which individuals are exposed \parencite{WHO08}. These dimensions include monitoring tobacco use and prevention policies (indicator M), protecting people from tobacco smoke (indicator P), offering help to quit tobacco use (indicator O), warning about the dangers of tobacco (indicator W), enforcing bans on tobacco advertising, promotion, and sponsorship (indicator E), and raising taxes on tobacco (indicator R). Each indicator ranges from one to four or five, with higher values indicating stricter regulations. Indicators P, O, W, and E are included as control variables. We exclude indicator M, as it reflects the monitoring of policies rather than a policy itself. Indicator R is also excluded because the tax share serves as the treatment in the tax-share analysis. In the price analysis, we do not adjust for price changes driven by taxes in order to assess the full impact of price changes. 

Considering treatment history allows us to compare individuals at similar pre-treatment price and tax-share levels. Including socio-demographic characteristics controls for potential confounding due to diverging demographic trends. We also account for other tobacco control policies to isolate the effect of price and tax-share increases, as such policies may be implemented simultaneously. \Cref{tab:meansOutCov} presents summary statistics of the control variables, reporting their means and standard deviations within the control and treatment groups.

\section{Method}
\label{sec:method}

In the main analysis, we apply the DiDDML estimator by \textcite{Zimmert20} to estimate the effect of price and tax-share increases on the smoking rates, controlling for treatment history, individuals' socio-demographic characteristics, and other tobacco prevention policies. Compared to conventional DiD approaches, DiDDML estimation implies a few modifications. First, outcome and treatment models are estimated in a data-driven way using machine learning. For this reason, treatment effect estimation is typically based on doubly robust score functions, see \textcite{Robins94}, which satisfy Neyman orthogonality. This implies that effect estimation is rather robust to approximation errors in the machine learning steps under certain regularity conditions. Furthermore, cross-fitting is applied to estimate the machine learning-based models for the treatment and outcome on the one hand, and the score function for effect estimation on the other, in different parts of the data. This avoids correlations between the estimation steps, a problem known as overfitting. When combining doubly robust score functions with cross-fitting, treatment effect estimation is asymptotically normal and root-n-consistent under certain regularity conditions despite the use of the additional machine learning steps. 

The essential differences between the DiDDML and parametric DiD approaches are that DiDDML requires that the score function is Neyman orthogonal and that cross-fitting is applied. Another key distinction is that the selected machine learning method for estimating outcome and treatment models must satisfy certain regularity conditions. The DiDDML estimator has multiple advantages over commonly used parametric DiD approaches. By using machine learning methods, we relax functional form assumptions, allowing for flexible control of high-dimensional confounding in a data-driven way. Moreover, the doubly robust score function allows for limited approximation errors in treatment and outcome models under certain regularity conditions.

\subsection{Identification Strategy}
\label{sec:idStrategy}

We use the potential outcomes framework, as described by \textcite{Rubin74}, to discuss the identification strategy. Let $D$ be a binary treatment, where $D = 0$ indicates no treatment and $D = 1$ indicates treatment. Let $T$ be a binary time period, where $T = 0$ is the pre-treatment period and $T = 1$ is the post-treatment period. The covariates $X$ represent observed factors. The observed outcomes are denoted by $Y_0$ in the pre-treatment period and $Y_1$ in the post-treatment period. The potential outcomes are $Y_t(0)$ if not treated and $Y_t(1)$ if treated in time period $t \in \{0, 1\}$. The treatment effect of interest, average treatment effect on the treated (ATET) in the post-treatment period ($T=1$), is defined as $\Delta_{D=1, T=1} = \mathbb{E}[Y_1(1) - Y_1(0) \mid D=1, T=1]$.

We aim to identify the ATET in the post-treatment period, based on the following five identification assumptions \parencite{Abadie05, Lechner11}: 
\begin{align}
	&Y_t = d Y_t(1) + (1-d) Y_t(0), \forall t \in \{0,1\}; X(1) = X(0) = X, \forall x \in X; \notag \\
	&\mathbb{E}[Y_0(1) - Y_0(0) \mid D=1, T=1, X] = 0; \notag \\
	&\mathbb{E}[Y_1(0) - Y_0(0) \mid D=0, X] = \mathbb{E}[Y_1(0) - Y_0(0) \mid D=1, X]; \notag \\
	&\Pr(D = 1, T = 1  \mid  (D, T) \in \{(d, t), (1, 1)\}, X) < 1, \forall (d, t) \in \{(1, 0), (0, 1), (0, 0)\}. \label{eq:idAssumptions}
\end{align}
First, the observation rule states that the observed outcome and the potential outcome conditional on treatment status are identical, and that the treatment assignment of any unit does not affect the potential outcomes of any other unit. Second, we assume exogeneity of the covariates, meaning that the treatment $D$ does not affect any of the covariates $X$. Third, the no anticipation assumption requires that the treatment assignment does not affect potential outcomes in the pre-treatment period. Fourth, the conditional common trend assumption states that the trends in the mean potential outcomes, given covariates $X$, are identical among the treated ($D=1$) and non-treated ($D=0$). Lastly, common support assumes that for any value of the covariates $X$ among the treated in the post-treatment period ($D=1, T=1$), the value must also occur in each other comparison group ($D=1, T=0$), ($D=0, T=1$), and ($D=0, T=0$).

\textcite{Zimmert20} shows that the ATET in the post-treatment period can be identified using a doubly robust score function that combines outcome regression and inverse probability weighting. Given the treatment $D$, time period $T$, and covariates $X$, the conditional mean outcome is defined as $\mu_D(T,X) = \mathbb{E}[Y \mid D,T,X]$ and the treatment propensity score is defined as $\rho_{D,T}(X) = \Pr(D,T \mid X)$, both referred to as nuisance parameters. Under the identification assumptions in equation (1), the ATET in the post-treatment period $\Delta_{D=1, T=1}$ is identified by:
\begin{align}
	\Delta_{D=1, T=1}&= \mathbb{E}\Bigg[ \Bigg\{ \frac{D\cdot T}{ \Pi} - \frac{D\cdot (1-T)\cdot \rho_{1, 1}(X)}{ \rho_{1, 0}(X)\cdot \Pi} \notag \\
	&\! - \Bigg(\frac{(1-D)\cdot T\cdot \rho_{1, 1}(X)}{ \rho_{0, 1}(X) \cdot \Pi} - \frac{(1-D)\cdot (1-T)\cdot \rho_{1, 1}(X)}{ \rho_{0, 0}(X)\cdot \Pi}\Bigg) \Bigg\} \times \Big(Y-\mu_D(T, X)\Big)  \notag \\
	&\! + \frac{D\cdot T}{ \Pi} \cdot \Bigg(\mu_1(1, X)-\mu_1(0, X)-\Big(\mu_0(1, X)-\mu_0(0, X)\Big) \Bigg) \Bigg],  \label{eq:scoreFunction}
\end{align}
where $\Pi=\Pr(D=1, T=1)$ is the unconditional probability of being in the post-treatment treated group. Note that in contrast to other DiD identification results which rely on the strong stationarity assumption (e.g., \textcite{Abadie05}, \textcite{SantAnna20}, and \textcite{Chang20}), this approach allows for varying distributions of covariates $X$ within treatment groups over time.

\subsection{Empirical Strategy}
\label{sec:empStrategy}

The variation in cigarette prices and taxes across countries and over time is exploited to estimate their effects on smoking. Individuals residing in countries that experience substantial increases in cigarette prices and in the tax share ($D = 1$) are compared with individuals in countries with stable price and tax environments ($D = 0$), following the treatment definitions provided in \cref{sec:dataTreatment}. Smoking outcomes in these two groups are compared using a DiD approach. 

We assess the price and tax-share increases between 2012--2014 and 2018--2020. For the 2012--2014 period, the pre-treatment period ($T = 0$) is 2012 and the post-treatment period ($T = 1$) is 2014, with an analogous definition for the 2018--2020 period. We pool the treatment effects across these two periods to make best use of the available data. Other time period combinations, such as 2014--2018, are not considered in order to pool only similar and thus comparable intervals.

The analysis focuses on the effect of large increases in prices (+15\%) and in the tax share (+2\%) between the periods 2012--2014 and 2018--2020. It is intuitive that large changes in prices and in the tax share are expected to be most effective, as also argued in \textcite{Callison14}. Changes in prices were selected because these are what final consumers face when purchasing a pack of cigarettes. However, identification of causal effects may be limited due to potential endogeneity, as discussed in more detail below. For this reason, changes in the tax share are appealing, as they are more independent of smoker behavior. Importantly, changes in absolute taxes were not considered because they still depend on the retail price of cigarette packs. Tax-share changes depend on tax policy rather than on price fluctuations, and may therefore be more credibly exogenous. For that reason, increases in the tax as a percentage of the retail price, i.e., the tax share, were selected as the tax treatment.

Cigarette prices may react to changes in cigarette demand in imperfect markets. The tobacco industry may respond to increased demand by raising prices. An increase in the smoking rate may therefore coincide with higher cigarette prices, biasing the estimated effect of price increases on smoking upward. Since the expected causal effect is negative, severe bias could even lead to sign reversals. This bias may be present in our study because pre-treatment smoking rates are higher among individuals experiencing a price increase than among those in the control group, as shown in \cref{tab:smokingRateInTreatmentGroups}. However, even if present, this bias would only attenuate the estimated effect size, as we expect smoking to decline in response to price increases.

One further endogeneity concern is reverse causality regarding tax-share increases. Governments may introduce additional tax increases in response to particularly high smoking rates. In this case, a rise in the smoking rate would coincide with tax increases, biasing the estimated effect of tax-share increases upward. Even if present, such bias would attenuate the estimated effect size. Unlike cigarette prices, the similar pre-treatment smoking rates across the treatment groups of the tax-share analysis suggest that reverse causality is not present, as shown in  \cref{tab:smokingRateInTreatmentGroups}. Bias from reverse causality therefore appears unlikely for the tax-share treatment, and even if present, it would only attenuate the estimated effect size.

Treatment history, as well as socio-demographic and policy covariates, are controlled for. Heterogeneous developments across treatment groups in these dimensions may lead to confounding. It is particularly important to control for other tobacco policies, such as smoking bans, health warnings, and advertising restrictions, because these may occur simultaneously with changes in cigarette prices or in the tax share. Moreover, heterogeneous demographic developments across treatment groups in determinants of smoking may also confound effect estimates. For this reason, gender, age, and proxies for socio-economic status are included as controls. Finally, controlling for treatment history, namely pre-treatment price and tax-share levels, ensures that price and tax-share increases are compared at similar levels.

The price and tax data do not allow us to verify parallel trends or to run placebo tests in the conventional way. Due to the high variation in prices and taxes, it is generally a challenge to run proper placebo tests using pre-treatment periods. Such tests would require a treatment definition under which, in at least two pre-treatment periods, neither the control nor the treatment group experiences a price or tax-share increase. As an alternative, we implement a related test suggested by \textcite{Callison14}, as described in \cref{sec:robustChecks}. 

Spillover effects may influence the analysis. \textcite{Bishop18} show that the effectiveness of cigarette tax increases in one geographic region depends on taxation in neighboring regions. Their study indicates that taxes are most effective in low-tax and low-price regions, where incentives for cross-border cigarette purchases are limited. In contrast, price and tax-share increases in high-tax and high-price regions may be less effective due to increased cross-border purchases, potentially biasing estimates toward zero. Consequently, our effect estimates may again represent a lower bound due to these possible spillover effects.

The survey data pose the limitation that smoking status was not always elicited consistently. The survey questions on smoking, namely the outcome variables, are identical in 2012, 2014, and 2020, but differ slightly in 2018. In 2012, 2014, and 2020, the questions refer to smoking boxed cigarettes and hand-rolled cigarettes, whereas in 2018 they refer to smoking traditional cigarettes.

The price measurements may pose limitations. In contrast to tax-share measures, cigarette prices are likely to vary within countries. Our price data represent prices in the country’s capital city \parencite{WHO22prices}, meaning that not all individuals in the survey necessarily face the same price. Moreover, \textcite{Pesko16} show that the estimated effect of prices varies with the calculation method, raising concerns about using price changes for impact evaluation. By comparison, tax-share changes are typically applied uniformly at the country level, resulting in more homogeneous exposure within countries. The tax-share measure therefore appears more reliable.

Finally, we cannot account for consumption of e-cigarettes and other emerging tobacco and nicotine products. If cigarettes and these new products are substitutes, effect estimation may be affected. Smoking may decline over time due to increased consumption of these emerging products. However, consumption rates remained relatively low over the time window of the analysis, with the share of e-cigarette users in OECD countries rising from 3\% in 2016 to 6\% in 2023 \parencite{OECD25}. In contrast to the previous limitations, the direction of this bias would not attenuate estimates toward zero and may instead lead to effect overestimation.

\subsection{Estimation}

For estimation, we apply the DiDDML estimator for repeated cross-sections by \textcite{Zimmert20}. The estimator is the sample analogue of the identification result in \cref{eq:scoreFunction}. Because the score function is doubly robust and Neyman orthogonal, the estimator is insensitive to moderate regularization bias in the machine learning estimation of the nuisance parameters. To avoid overfitting bias, cross-fitting is applied such that the nuisance parameters and the treatment effect are estimated in different parts of the data. The ATET in the post-treatment period is estimated based on the following cross-fitting algorithm:
\begin{enumerate} 
	\item Randomly split the sample into $k \in \{1, 2, ..., K\}$ non-overlapping and equally sized subsamples, also called folds.
	\item Estimate the models of the nuisance parameters, $\hat{\mu}_D(T,X)$ and $\hat{\rho}_{D,T}(X)$ for all $d,t \in \{0, 1\}$, using the whole sample except fold $k$.
	\item Predict the nuisance parameters based on the previously estimated models, plug them into the score function in \cref{eq:scoreFunction}, and estimate the treatment effect, using the remaining fold $k$.
	\item Repeat steps 2 and 3 $K$ times, once for each fold.
	\item Averaging the treatment effects yields the ATET in the post-treatment period.
\end{enumerate}
For the main analysis, standard errors are clustered at the country level. In step 2, the nuisance parameters are estimated using various machine learning methods, such as lasso and random forests. For our main analysis, we use random forests, as discussed in \textcite{Breiman01}. Building on decision trees that predict the treatment or outcome based on recursively splitting the data into subsets according to predictive values of the covariates, random forests additionally rely on subsampling and random feature selection at splitting decisions. The advantage of random forests, compared to other machine learning methods, is that they are nonparametric and thus do not require strong functional form assumptions. Under the regularity condition that the random forest-based approximations of the nuisance parameters converge at a rate of $o(N^{-1/4})$, the DiDDML estimator is root-n-consistent and asymptotically normal for estimating the ATET in the post-treatment period.

The main analysis estimates the impact of price and tax-share increases on smoking rates among current and daily smokers. We implement the DiDDML estimator with random forests by \textcite{Zimmert20}, controlling for all covariates, including treatment history, individuals’ socio-demographic characteristics, and tobacco prevention policies. Results are pooled across the periods 2012--2014 and 2018--2020, with standard errors clustered at the country level. We enforce common support by trimming extreme treatment propensity scores in the respective comparison groups (other than the reference group of post-treatment treated observations) below 0.01, following \textcite{Crump09}. Ten folds are used for $k$-fold cross-fitting. The analysis is implemented in R using the function \texttt{didDML} from the \texttt{causalweight} package by \textcite{Bodory24}, which is based on the methodology introduced by \textcite{Zimmert20}.

In contrast to conventional TWFE models, we do not include region fixed effects, such as country fixed effects. Because treatments are defined at the country level and only a few time periods are observed, including country fixed effects would produce extreme propensity scores and lead to unstable estimation. We also stack pre- and post-treatment periods across analysis periods, coding 2012 and 2018 as zero and 2014 and 2020 as one. In addition, we include a dummy variable indicating the analysis period. This approach avoids the “negative weights” bias that may arise in settings with variation in treatment timing.

For comparison, we additionally apply a parametric DiD estimator. Following a conventional TWFE specification, the outcome is regressed on a constant, the treatment, covariates, and year and country fixed effects. Given individuals $i$ in countries $j$ and years $t$, the treatment effect is estimated based on the following regression model:
\begin{align}
	Y_{ijt}	= \alpha + \theta D_{jt} + \sum_{k=1}^{K} \beta_{X_k} X_{itk}
	+ \gamma_j + \tau_t + \epsilon_{ijt}, \label{eq:paramDiD}
\end{align}
where $Y_{ijt}$ is the outcome, $\alpha$ is a constant, $\theta$ captures the marginal effect of the treatment $D_{jt}$, $X_{itk}$ are covariates, $\gamma_j$ are country fixed effects, $\tau_t$ are year fixed effects, and $\epsilon_{ijt}$ is an idiosyncratic error term. Standard errors are clustered at the country level.

In this parametric DiD approach, the treatment $D_{jt}$ may be specified as either binary or continuous. We provide estimates from a parametric DiD model with binary treatments, which is closest to the DiDDML estimates. We also implement a parametric DiD model with continuous treatments. Instead of indicating large price or tax-share increases, the continuous treatment corresponds to the absolute cigarette price or tax share. This specification is informative because it is typically used in analyses of tobacco prices and taxes, as discussed in \cref{sec:literature}. It estimates the marginal effect of a one-unit increase in price or in the tax share, which also makes it conventional to directly estimate elasticities. While both parametric DiD approaches are technically implementations of TWFE, the latter version with continuous treatments is the one usually applied and thus commonly referred to as TWFE in the literature on tobacco prices and taxes.

In comparison to DiDDML, the parametric DiD approach relies on stronger functional form assumptions, such as covariates entering the outcome additively and linearly. In addition, parametric DiD with continuous treatments imposes homogeneous treatment effects across all treatment doses and values. It is also important to note that estimates based on binary and continuous treatments are not directly comparable. The continuous-treatment specification estimates marginal effects of one-unit increases in prices and tax shares, whereas the binary-treatment specification compares outcomes of well-defined control and treatment groups.

\section{Results}
\label{sec:results}

The effects of price and tax-share increases on smoking are discussed in the following. We first present the estimated effects of large price and tax-share increases and the resulting price elasticities of cigarette demand. Moreover, estimates based on flexible and the parametric DiD approaches are compared. Next, we examine the sensitivity of these results to the treatment definition, comparing binary and continuous treatments. Finally, heterogeneous effects by gender, age, and education are discussed.

\subsection{Effect of Price and Tax Increases}

\begin{figure}[htbp!]
	\centering
	\includegraphics[width=0.8\textwidth]{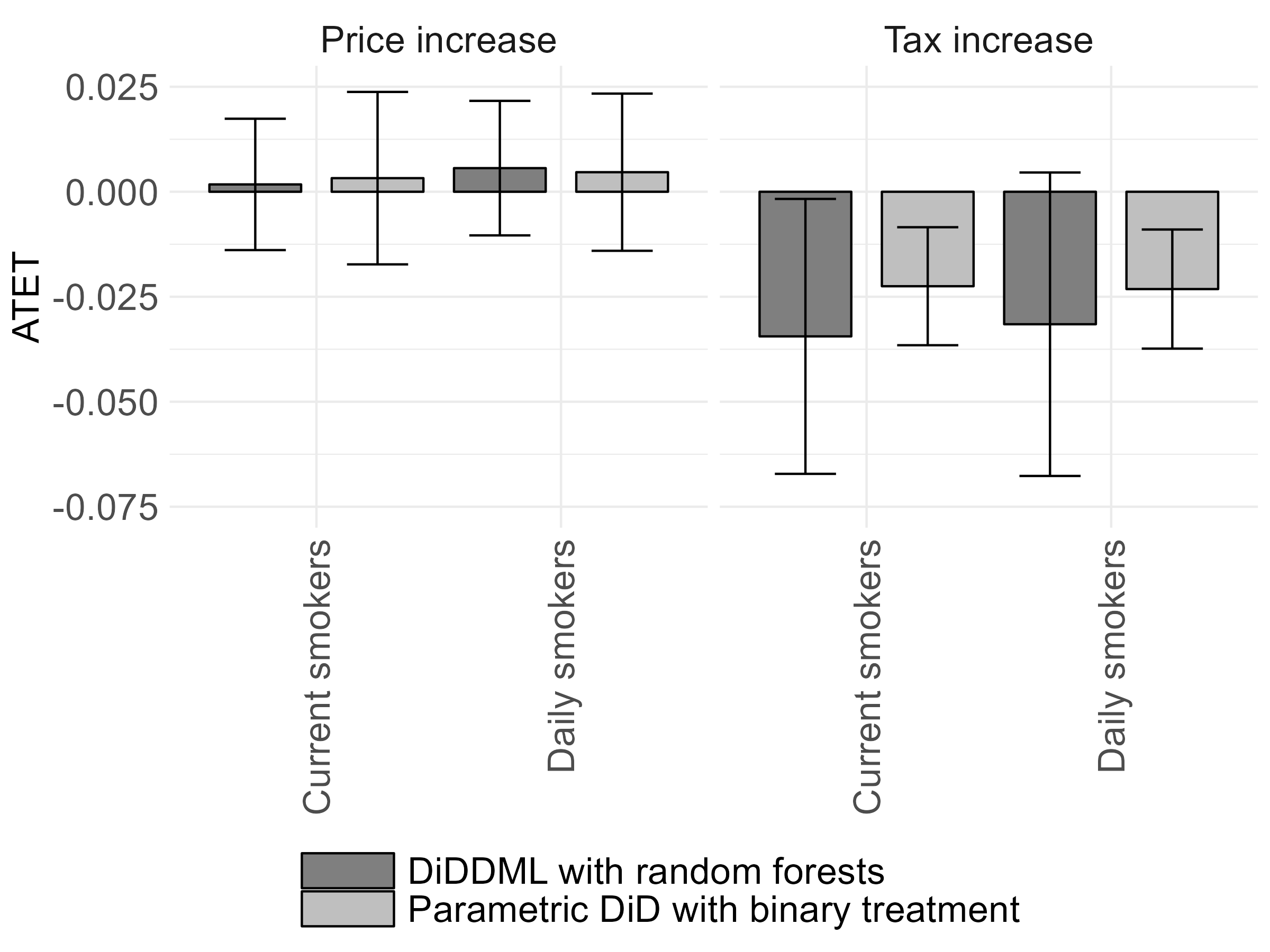}
	\caption{Effect of price and tax increases on the smoking rate}
	\label{fig:mainRes}
	\caption*{\footnotesize \textit{Note: This figure displays the effect of price and tax increases on the smoking rate, corresponding to the estimates reported in columns (4) and (6) of \cref{tab:mainRes}. The results are based on DiDDML with random forests and on the parametric DiD with a binary treatment, as defined in \cref{eq:paramDiD}. Error bars represent 95\% confidence intervals.}}
\end{figure}

\Cref{fig:mainRes} shows the effect of a price increase and of a tax-share increase on the smoking rates of both current and daily smokers. For simplicity, we refer to an increase in the tax-share simply as a tax increase in the remainder of this paper. The main results are the DiDDML estimates, corresponding to column (4) in \cref{tab:mainRes}. In addition, estimates based on the parametric DiD model with binary treatments, specified in \cref{eq:paramDiD}, are presented and correspond to column (6) in \cref{tab:mainRes}.

Regarding price increases, the DiDDML estimates show that the post-treatment ATETs are close to zero and not statistically significant at the 5\% level for both current and daily smoking rates. The parametric DiD estimates yield similarly small and statistically insignificant effects. As discussed in \cref{sec:empStrategy}, the estimates may be biased upward due to endogeneity, as price increases may themselves be driven by smoking demand. Consequently, this result should be interpreted with caution, and the null findings do not rule out that prices may still affect smoking.

Regarding tax increases, the DiDDML estimates show reductions in post-treatment smoking rates of current smokers by 3.44 pp (a 15\% reduction; p $=$ 0.04) and of daily smokers by 3.15 pp (a 15\% reduction; p $=$ 0.09). The parametric DiD estimates also indicate negative and statistically significant effects for both groups, with current smokers experiencing a reduction of 2.25 pp (a 10\% reduction; p $<$ 0.01) and daily smokers a reduction of 2.32 pp (a 11\% reduction; p $<$ 0.01). Overall, these findings indicate that increases in the tax share reduce smoking rates among both current and daily smokers.

We verify the common support assumption, which requires that treatment is not deterministic. Accordingly, for each covariate composition among treated individuals in the post-treatment period ($D = 1, T = 1$), there must be at least one individual in each of the remaining comparison groups ($D = 1, T = 0$), ($D = 0, T = 1$), and ($D = 0, T = 0$). \Cref{fig:propensityScores} plots the treatment propensity scores of all four groups, based on random forests. Common support holds because the propensity-score distribution of the treated in the post-treatment period overlaps with the distributions of the comparison groups.

The DiDDML estimates, which allow for flexible confounding, and the parametric DiD estimates yield qualitatively similar results. Both approaches show null effects for the price increase and negative effects for the tax increase. While the estimates for price increases are very similar, the differences for tax increases are more pronounced. Compared with DiDDML, the parametric DiD estimates are roughly one-fourth smaller and have much narrower confidence intervals, producing p-values below 0.01. However, the confidence intervals largely overlap, indicating that the DiDDML and parametric DiD estimates are qualitatively similar. Overall, treatment effect estimates do not appear to be particularly sensitive to these functional form assumptions. 

The pass-through rate from the tax increase to prices is estimated by replacing the smoking outcome with cigarette prices in the parametric DiD specification with the binary tax treatment. The tax increase leads to a price increase of 0.22 PPP\$, given a post-treatment average price of 8.65 PPP\$ among the treatment groups. The average tax increase of 4.13\% among the treatment groups corresponds to a 0.29 PPP\$ increase in prices. Consequently, the pass-through rate is 77\% (equal to 0.22 PPP\$ divided by 0.29 PPP\$). This is in line with the literature, as most studies find that a substantial share of cigarette taxes is passed on to consumers through higher retail prices \parencite{DeCicca22}. The assessed tax increases therefore appear to translate into higher prices, through which they ultimately influence smoking behavior.

Focusing on the tax increases, we may derive the price elasticity of cigarette demand at the extensive margin. The elasticity is computed as the percentage change in the smoking rate due to the tax increase divided by the tax-induced price increase in percent. This results in elasticities of -5.7 and -5.6 based on the DiDDML estimates, while the parametric DiD estimates yield elasticities of -3.9 and -4.3. These values are multiple times larger than commonly estimated extensive-margin price elasticities, which typically range between -0.1 and -0.3 \parencite{DeCicca22}. However, our estimates are based on comparisons of individuals exposed to large tax increases with individuals without tax changes. Hence, these estimates, which rely on a DiD design with well-defined control and treatment groups, may not be directly comparable to existing elasticity estimates based on parametric DiD specifications that include prices and taxes as continuous regressors. Another factor may be the emergence of new tobacco and nicotine products during 2012--2020, which could increase substitution. 

While the point estimates are large compared to the literature, the confidence intervals are relatively wide. Considering the lower bound, conservative estimates become -0.3 for current smokers based on the DiDDML estimates. For daily smokers, the confidence interval even includes zero. The conservative elasticities based on the parametric DiD approach remain larger in magnitude. These estimates become -1.6 for current smokers and -1.8 for daily smokers based on the parametric DiD estimates. Overall, our findings indicate that the extensive-margin price elasticity of cigarette demand is negative, and that its magnitude might be larger than current elasticity estimates suggest.

\FloatBarrier

\subsection{Binary Versus Continuous Treatment Definitions}

Both main estimates above rely on binary treatments, once based on the flexible DiDDML estimator and once based on the parametric DiD estimator. This is our preferred specification, as the binary treatment allows for a clean comparison of individuals exposed to a large price and tax increase with individuals experiencing stable prices and taxes. However, the literature on tobacco prices and taxes typically relies on parametric DiD using continuous treatments. For that reason, the sensitivity of the estimates to varying treatment definitions will be explored in the following.

The parametric DiD estimator using binary treatments in \cref{eq:paramDiD} is first applied to the sample with large price and tax increases. Second, the same parametric DiD estimator is applied using continuous treatments, i.e., the absolute value of prices and taxes, instead of binary treatments. This estimate is still based on the restricted sample with large increases. Third, the continuous-treatment specification is applied to the full sample, including weak increases as well as all decreases in prices and taxes. The latter specification is the most common approach in the literature, as discussed in \cref{sec:literature}.

Columns (6), (7), and (8) in \cref{tab:mainRes} report the resulting estimates. Column (6) coincides with the previously discussed main results, as this estimate is based on the parametric DiD estimation with the binary treatment applied to the sample restricted to large increases. Column (7) then reports results based on the continuous-treatment specification applied to the same restricted sample, and column (8) shows the results applying it to the full sample. It is important to note that the latter two estimates correspond to the marginal effect of a one-unit increase in the price and tax treatment, and are therefore not directly comparable to the estimates based on binary treatments.

\begin{figure}[htbp!]
	\centering
	\includegraphics[width=0.8\textwidth]{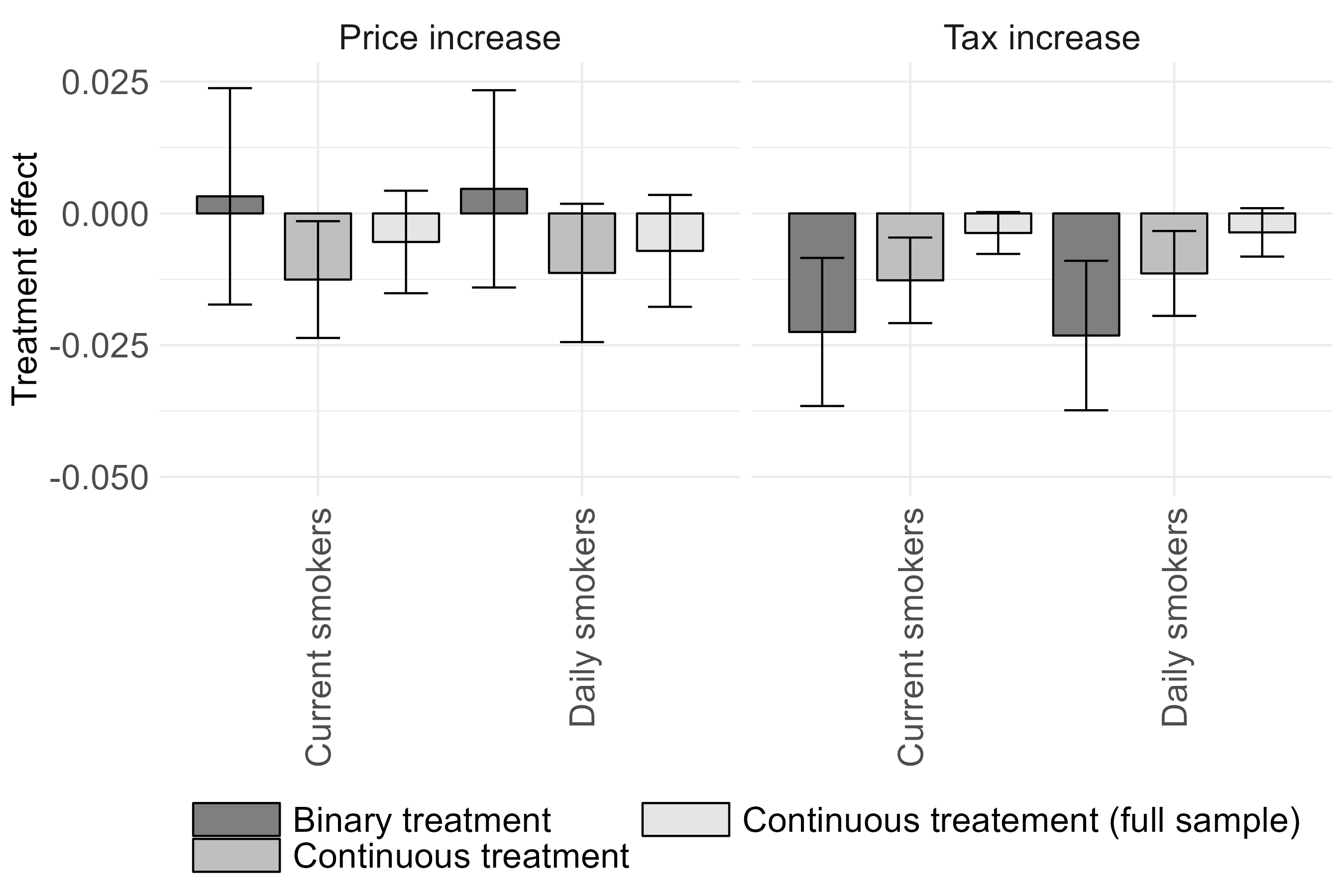}
	\caption{Comparison of parametric DiD estimates with varying treatment definitions}
	\label{fig:mainResComparison}
	\caption*{\footnotesize \textit{Note: This figure shows the effect of price and tax increases on the smoking rate. It compares different treatment definitions using the parametric DiD approach, namely column (6) with a binary treatment in large price- and tax-increase sample, column (7) with a continuous treatment in large price- and tax-increase sample, and column (8) with a continuous treatment in the full sample, as presented in \cref{tab:mainRes}. Error bars represent 95\% confidence intervals.}}
\end{figure}

\Cref{fig:mainResComparison} compares the parametric DiD estimates based on binary treatments with the estimates based on continuous treatments. To ensure comparability, the estimates using continuous treatments are rescaled to reflect the expected effect of a 27.20\% price increase and a 4.13\% tax increase, as in the binary-treatment estimates. In the following, all reported effect sizes and significance levels correspond to these rescaled and thus comparable estimates.

Regarding price increases, the estimates based on binary treatments were close to zero, yielding null effects. In contrast, the estimates based on continuous treatments indicate reductions in smoking rates and are larger in magnitude. Based on the restricted sample with large price increases, the estimates with continuous treatments suggest reductions of 1.26 pp (p $=$ 0.03) in the current smoking rate and 1.13 pp (p $=$ 0.09) in the daily smoking rate. These negative effects become smaller in the full sample, namely reductions of 0.54 pp (p $=$ 0.27) among current smokers and 0.71 pp (p $=$ 0.19) among daily smokers.

Regarding tax increases, both the estimates based on binary and continuous treatments point to reductions in smoking rates. However, the estimates based on continuous treatments are considerably smaller. Based on the restricted sample with large tax increases, the estimates based on continuous treatments suggest reductions of 1.27 pp (p $<$ 0.01) among current and 1.14 pp (p $<$ 0.01) among the daily smokers. This is less than half the effect size of the binary-treatment estimates, which find reductions of 2.25 pp and 2.32, respectively. The estimates based on continuous treatments in the full sample are even smaller, with reductions of 0.37 pp (p $=$ 0.07) among current smokers and 0.36 pp (p $=$ 0.12) among daily smokers.

The comparison shows that effect estimates are sensitive to the treatment definition. Estimates using binary treatments yield larger reductions in smoking rates compared to the parametric estimates using continuous treatments. This may help explain why our main estimates of elasticities of cigarette demand differ from the existing literature. Moreover, estimates based on the sample with large price and tax increases are larger in magnitude than those based on the full sample. Because the binary DiD design provides a clearer counterfactual by comparing well-defined treatment and control groups, elasticity estimates based on continuous treatments that rely on strong linearity assumptions and the full sample may have been underestimated. This suggests that the linearity assumption imposed by the parametric DiD model with continuous treatments may not hold.

\FloatBarrier

\subsection{Effect Heterogeneity}

We assess the effect heterogeneity of price and tax increases across different subgroups based on gender, age, and education. We apply the same model and parameters as in the main analysis to these subgroups, additionally accounting for multiple hypothesis testing with the Benjamini-Hochberg method \parencite{Benjamini95}. This method controls the false discovery rate, i.e., the proportion of incorrectly rejected null hypotheses among all the rejected null hypotheses. We evaluate the effects separately for men and women, as well as for the age groups 15--24, 25--44, 45--64, and 65+. Additionally, we consider individuals who stop full-time education at different ages: up to 15 years old, 16--19 years old, and 20+ years old. The age at which one stops full-time education serves as a proxy for the level of education, corresponding to low, middle, and high levels of education, respectively.

\Cref{tab:hetRes} presents the heterogeneous effects of price and tax increases on smoking rates. The estimates indicate that price increases have no effect on smoking rates in either the main analysis or any subgroup. In contrast, the tax-increase estimates show reductions in smoking rates across all subgroups. The effect of tax increases is statistically significant at the 5\% level for both current and daily smokers aged 15--24. In addition, women, individuals aged 45--64, and those with low or medium levels of education show statistically significant effects at the 10\% level among current smokers, though not among daily smokers. Overall, the reduction in smoking rates due to tax increases is mainly driven by individuals aged 15--24.

Our results align with studies showing that young individuals are particularly affected by tax increases, such as \textcite{Odermatt18} and \textcite{Pesko16}. This finding is theoretically sound, as young individuals have relatively little available income and are therefore expected to respond more strongly to price and tax mechanisms. While e-cigarette uptake in the general population is limited, uptake is relatively high among youth. Hence, part of this heterogeneous effect may be explained by greater opportunities for substitution among younger individuals. 

Regarding gender, we find that reductions in smoking rates are stronger among women, consistent with \textcite{Maclean14} and \textcite{Pesko20}, but in contrast to \textcite{Odermatt18}.

However, we find slightly more pronounced effects among individuals with low or middle levels of education compared to those with high levels of education, i.e., among individuals with relatively lower socio-economic status. In contrast, \textcite{Maclean14} and \textcite{Pesko16} report that highly educated individuals react particularly strongly to tax increases. Our findings appear more intuitive, as individuals with low and middle levels of education likely have less available income and therefore should respond more strongly to price and tax mechanisms.

\section{Robustness Checks}
\label{sec:robustChecks}

The conditional common trend assumption cannot be tested using placebo tests due to significant variation in prices and taxes over time. This is a common limitation in the literature on cigarette prices and taxes. To address this, we perform a placebo analysis inspired by \textcite{Callison14}. In this analysis, we only include control observations. First, we assign a placebo treatment to all observations within a control unit, which in this case is one country in either 2012--2014 or 2018--2020. Second, we calculate the placebo treatment effect based on the main analysis specifications, using the remaining control observations as the control group. These steps are repeated for each control unit. Finally, we average the placebo effects. The placebo analysis therefore tests whether the estimated placebo treatment effects are close to zero when no actual treatment has occurred. For price increases, the average placebo estimates are -0.2 pp (p $=$ 0.95) and -0.6 pp (p $=$ 0.86), for current and daily smokers, respectively. For tax increases, the average placebo estimates are close to zero for both current and daily smokers, -0.1 pp (p $=$ 0.98) and -0.3 pp (p $=$ 0.94), respectively. For all analyses, the average placebo effect is not significant at the 5\% level, supporting the validity of the DiD design. \Cref{fig:placeboTests} shows the distribution of the placebo effects for each of the four main analyses.

We test the robustness of our main specification to changes in the included covariates $X$. To do this, we perform the main analysis using three different sets of covariates: (1) treatment history, (2) treatment history and socio-demographic covariates, and (3) treatment history and policy covariates. Columns (1), (2), and (3) in \cref{tab:mainRes} show the corresponding estimates. For price increases, the effects remain close to zero and are not significant at the 5\% level, consistent with the main estimate in all three specifications. For tax increases, the effect direction is consistently negative and in multiple instances statistically significant at the 5\% level, also aligning with the main estimate in all three specifications. Therefore, the main analysis results appear to be robust to changes in covariate composition.

We cluster standard errors at the country level. Due to the limited number of countries, the resulting standard errors may be too conservative. However, considering the potential clustering of institutional and cultural factors within the same country, cluster-robust standard errors might be more appropriate in this context. Column (5) in \cref{tab:mainRes} shows the estimates using the main specification without clustering. For price increases, the effects remain insignificant regardless of clustering. For tax increases, both effects on current and daily smokers are significant at the 5\% level when clustering is not applied. When clustering is applied, the effect on current smokers remains significant at the 5\% level, while the effect on daily smokers remains significant at the 10\% level. Therefore, the main analysis results appear robust to changes in standard error clustering.

Our analysis includes one survey wave from 2020, which was collected during the COVID-19 pandemic. Countries may differ in their responses to the pandemic, introducing potential confounding factors such as the closure of restaurants and other social restrictions. To investigate this limitation, we perform the main analysis separately for the periods 2012--2014 and 2018--2020. \Cref{tab:mainRes12141820} presents the results of these analyses, which are consistent with the pooled main results. For tax increases, the effects consistently point in the same direction as the pooled results, although they are not significant at the 5\% level. For price increases, the effects remain statistically insignificant. These findings suggest that the pooled main results are robust across different time periods, indicating that the impact of the COVID-19 pandemic is limited.

\section{Conclusion}
\label{sec:conclusion}

Our study shows that cigarette tax increases reduce smoking rates. Using a DiD approach, we compare a group experiencing a large increase in the tax share with a group experiencing no change. The tax share is defined as the percentage of the retail cigarette price. The exposed group faced an increase of 4.13\% (0.29 PPP\$), such that approximately 77\% of this increase passed through to higher cigarette prices. Our estimates indicate that tax increases reduce the smoking rate by 3.44 pp (a 15\% reduction; p $=$ 0.04) for individuals who smoke at least once per month and by 3.15 pp (a 15\% reduction; p $=$ 0.09) for daily smokers among post-treatment individuals who experience the increase.

These results imply a price elasticity of cigarette demand at the extensive margin that is a multiple of existing estimates. However, these point estimates are rather imprecise, and the confidence intervals are relatively wide. Considering the lower bounds, the conservative elasticities become -0.3 for current smokers based on the DiDDML estimates, while for daily smokers the confidence interval even includes zero. In contrast, conservative elasticities based on the parametric DiD approach using binary treatment variables remain larger in magnitude, namely -1.6 for current smokers and -1.8 for daily smokers. Overall, these conservative estimates indicate that the extensive-margin price elasticity of cigarette demand is negative, and may exceed existing estimates, which range between -0.1 and -0.3 \parencite{DeCicca22}.

Our European evidence is consistent with findings from the United States, which also show that tax increases reduce smoking. Contrary to studies arguing that the effectiveness of tax increases diminishes or disappears over time, our analysis indicates that tax increases continue to reduce smoking rates. This remains true even in high-tax settings, where around three-fourths of the retail price are taxes, and during recent periods between 2012 and 2020. Eventually, the increased availability of e-cigarettes and thus substitution possibilities may also have contributed to this result.

Effect heterogeneity analysis shows reductions across all subpopulations by gender, age, and education level. But only the reduction among youth is statistically significant at the 5\% level. This suggests that the effect of tax increases is primarily driven by individuals aged 15--24. This finding aligns with evidence from the United States \parencite{Pesko16} and Europe \parencite{Odermatt18}. Youth have limited income and therefore appear to respond most strongly to price and tax mechanisms. Moreover, because e-cigarette uptake is most prevalent among youth, they may be more likely to substitute toward alternative nicotine and tobacco products than other groups.

Our analysis of price increases does not identify a statistically significant effect on smoking rates at the 5\% level. In our sample, large price increases occur in populations with relatively high smoking rates. This suggests that the industry may raise prices in response to elevated smoking rates. Consequently, these estimates may suffer from upward bias toward zero due to the endogeneity of price changes.

We examine the sensitivity of our estimates to functional form assumptions and alternative treatment definitions. While relaxing functional form assumptions does not appear to substantially affect the estimates, treatment definitions play an important role in estimating price and tax effects. Regarding the tax-increase effect, continuous-treatment specifications, such as TWFE models, may underestimate the impact of tax increases on smoking. This attenuation becomes even stronger when using the full sample instead of focusing on a restricted sample that includes only large tax increases and stable tax environments. In conclusion, our findings indicate that estimates based on continuous-treatment specifications and the full sample may underestimate the effectiveness of tobacco tax increases, compared to a DiD approach with well-defined control and treatment groups using a restricted sample.

\section*{Acknowledgments}

Thanks are expressed to Elsa Gautrain and Matthias Westphal, as well as participants at the Ski and Labor Economics Seminar, Symposium of Causal Inference in the Health Sciences, EuHEA PhD Student \& Supervisor Conference, Swiss Public Health Conference, \#PopHealthLab Science Meeting at the Population Health Laboratory, and internal seminar of our Department of Economics, for their valuable comments. Edwin Froidevaux is thanked for his research assistance. Financial support from the Swiss Tobacco Prevention Fund is gratefully acknowledged.

\section*{Declaration of Generative AI and AI-assisted Technologies in the Writing Process}
% https://www.elsevier.com/about/policies-and-standards/generative-ai-policies-for-journals

During the preparation of this work the authors used Microsoft Copilot in order to rephrase or partially generate text, as well as to refine and partially generate code. After using this tool, the authors reviewed and edited the content as needed and take full responsibility for the content of the publication.

% references

\section*{References}

\printbibliography[heading=none]

% appendix

\newpage

\begin{appendices}
	
\section{Annex}

% table treatments

\begin{table}[htbp!]
	\caption{Prices, changes in prices, and treatment assignment}
	\label{tab:treatmentsPrices}
	\centering
	\scriptsize
	\begin{tabular}{p{.15\textwidth}p{.08\textwidth}p{.08\textwidth}p{.08\textwidth}p{.08\textwidth}p{.08\textwidth}p{.08\textwidth}p{.08\textwidth}p{.08\textwidth}}
		& \multicolumn{4}{c}{2012--2014} & \multicolumn{4}{c}{2018--2020} \\
		\cmidrule(lr){2-5}\cmidrule(lr){6-9}
		Country & 2012 & 2014 & $\Delta$, in \% & $D$ & 2018 & 2020 & $\Delta$, in \% & $D$ \\ 
		\hline
		Austria & 6.15 & 6.65 & 8.10 &  & 7.40 & 7.42 & 0.22 & 0 \\ 
		Belgium & 7.12 & 7.86 & 10.32 &  & 8.88 & 8.89 & 0.15 & 0 \\ 
		Bulgaria & 7.39 & 7.23 & -2.19 & 0 & 7.22 & 7.06 & -2.20 & 0 \\ 
		Cyprus & 5.90 & 6.38 & 8.19 &  & 7.61 & 7.67 & 0.79 & 0 \\ 
		Czech Republic & 5.69 & 6.15 & 8.21 &  & 7.81 & 8.35 & 6.97 &  \\ 
		Denmark & 5.89 & 6.51 & 10.61 &  & 6.79 & 9.20 & 35.57 & 1 \\ 
		Estonia & 6.62 & 7.21 & 8.83 &  & 6.64 & 7.37 & 10.96 &  \\ 
		Finland & 6.00 & 6.54 & 9.10 &  & 8.66 & 10.07 & 16.28 & 1 \\ 
		France & 8.18 & 9.40 & 14.90 &  & 10.89 & 13.23 & 21.43 & 1 \\ 
		Germany & 7.45 & 7.73 & 3.79 & 0 & 8.97 & 9.43 & 5.14 &  \\ 
		Greece & 6.01 & 7.11 & 18.29 & 1 & 8.37 & 8.50 & 1.53 & 0 \\ 
		Hungary & 6.71 & 8.39 & 25.01 & 1 & 9.35 & 10.38 & 11.01 &  \\ 
		Ireland & 12.31 & 12.72 & 3.34 & 0 & 16.21 & 17.31 & 6.80 &  \\ 
		Italy & 7.43 & 7.34 & -1.31 & 0 & 8.38 & 8.82 & 5.22 &  \\ 
		Latvia & 3.96 & 6.53 & 64.91 & 1 & 7.32 & 7.51 & 2.57 & 0 \\ 
		Lithuania & 5.41 & 5.80 & 7.15 &  & 8.66 & 9.20 & 6.23 &  \\ 
		Luxembourg & 5.64 & 6.14 & 8.87 &  & 6.46 & 6.36 & -1.50 & 0 \\ 
		Malta & 8.06 & 8.89 & 10.32 &  & 9.83 & 9.52 & -3.20 & 0 \\
		Netherlands & 7.68 & 8.48 & 10.38 &  & 9.26 & 10.18 & 9.96 &  \\ 
		Poland & 7.19 & 8.41 & 16.99 & 1 & 9.30 & 9.29 & -0.09 & 0 \\ 
		Portugal & 7.72 & 8.43 & 9.18 &  & 9.03 & 7.71 & -14.63 &  \\ 
		Romania & 9.60 & 9.67 & 0.68 & 0 & 10.82 & 11.41 & 5.43 &  \\ 
		Slovakia & 6.03 & 6.36 & 5.44 &  & 6.79 & 6.90 & 1.68 & 0 \\ 
		Slovenia & 5.32 & 6.34 & 19.15 & 1 & 6.72 & 6.68 & -0.66 & 0 \\ 
		Spain & 7.20 & 7.87 & 9.28 &  & 8.28 & 8.13 & -1.80 & 0 \\ 
		Sweden & 6.81 & 7.33 & 7.56 &  & 7.67 & 7.71 & 0.50 & 0 \\ 
		United Kingdom & 10.46 & 11.69 & 11.73 &  & 14.23 & 14.91 & 4.77 & 0 \\
		\hline\hline
	\end{tabular}
	\caption*{\footnotesize \textit{Note: This table shows the pre- and post-treatment prices, price changes from pre- to post-treatment period, and treatment assignment within the periods 2012--2014 and 2018--2020 by country.}}
\end{table}

\newpage

\begin{table}[htbp!]
	\caption{Tax shares, changes in tax shares, and treatment assignment}
	\label{tab:treatmentsTaxes}
	\centering
	\scriptsize
	\begin{tabular}{p{.15\textwidth}p{.08\textwidth}p{.08\textwidth}p{.08\textwidth}p{.08\textwidth}p{.08\textwidth}p{.08\textwidth}p{.08\textwidth}p{.08\textwidth}}
		& \multicolumn{4}{c}{2012--2014} & \multicolumn{4}{c}{2018--2020} \\
		\cmidrule(lr){2-5}\cmidrule(lr){6-9}
		Country & 2012 & 2014 & $\Delta$, in \% & $D$ & 2018 & 2020 & $\Delta$, in \% & $D$ \\ 
	\hline
		Austria & 0.74 & 0.74 & 0.00 & 0 & 0.75 & 0.75 & 0.00 & 0 \\ 
		Belgium & 0.76 & 0.76 & 0.00 & 0 & 0.77 & 0.77 & 0.00 & 0 \\ 
		Bulgaria & 0.84 & 0.86 & 2.38 & 1 & 0.87 & 0.85 & -2.30 &  \\ 
		Cyprus & 0.76 & 0.77 & 1.32 &  & 0.74 & 0.74 & 0.00 & 0 \\ 
		Czech Republic & 0.78 & 0.77 & -1.28 &  & 0.75 & 0.77 & 2.67 & 1 \\ 
		Denmark & 0.79 & 0.75 & -5.06 &  & 0.74 & 0.78 & 5.41 & 1 \\ 
		Estonia & 0.77 & 0.77 & 0.00 & 0 & 0.86 & 0.88 & 2.33 & 1 \\ 
		Finland & 0.80 & 0.82 & 2.50 & 1 & 0.87 & 0.88 & 1.15 &  \\ 
		France & 0.80 & 0.80 & 0.00 & 0 & 0.82 & 0.83 & 1.22 &  \\ 
		Germany & 0.73 & 0.73 & 0.00 & 0 & 0.68 & 0.64 & -5.88 &  \\ 
		Greece & 0.82 & 0.80 & -2.44 &  & 0.81 & 0.81 & 0.00 & 0 \\ 
		Hungary & 0.84 & 0.77 & -8.33 &  & 0.72 & 0.73 & 1.39 &  \\ 
		Ireland & 0.79 & 0.78 & -1.27 &  & 0.78 & 0.79 & 1.28 &  \\ 
		Italy & 0.75 & 0.76 & 1.33 &  & 0.76 & 0.77 & 1.32 &  \\ 
		Latvia & 0.79 & 0.77 & -2.53 &  & 0.80 & 0.80 & 0.00 & 0 \\ 
		Lithuania & 0.75 & 0.76 & 1.33 &  & 0.74 & 0.74 & 0.00 & 0 \\ 
		Luxembourg & 0.71 & 0.70 & -1.41 &  & 0.68 & 0.68 & 0.00 & 0 \\ 
		Malta & 0.77 & 0.75 & -2.60 &  & 0.78 & 0.78 & 0.00 & 0 \\ 
		Netherlands & 0.72 & 0.73 & 1.39 &  & 0.72 & 0.77 & 6.94 & 1 \\ 
		Poland & 0.80 & 0.80 & 0.00 & 0 & 0.77 & 0.78 & 1.30 &  \\ 
		Portugal & 0.76 & 0.75 & -1.32 &  & 0.72 & 0.79 & 9.72 & 1 \\ 
		Romania & 0.73 & 0.75 & 2.74 & 1 & 0.69 & 0.70 & 1.45 &  \\ 
		Slovakia & 0.82 & 0.82 & 0.00 & 0 & 0.77 & 0.76 & -1.30 &  \\ 
		Slovenia & 0.79 & 0.80 & 1.27 &  & 0.79 & 0.79 & 0.00 & 0 \\ 
		Spain & 0.79 & 0.78 & -1.27 &  & 0.78 & 0.78 & 0.00 & 0 \\ 
		Sweden & 0.74 & 0.69 & -6.76 &  & 0.68 & 0.68 & 0.00 & 0 \\ 
		United Kingdom & 0.80 & 0.82 & 2.50 & 1 & 0.79 & 0.79 & 0.00 & 0 \\  
		\hline\hline
	\end{tabular}
	\caption*{\footnotesize \textit{Note: This table shows the pre- and post-treatment tax shares, tax share changes from pre- to post-treatment period, and treatment assignment within the periods 2012--2014 and 2018--2020 by country.}}
\end{table}

\newpage

% table means

\setcounter{cTable}{\value{table}}
\refstepcounter{cTable}
\label{tab:meansOutCov}
\scriptsize
\setlength\LTleft{0pt}
\setlength\LTright{0pt}
\begin{longtable}{p{.35\textwidth}p{.15\textwidth}p{.15\textwidth}p{.15\textwidth}p{.15\textwidth}}
	\caption{Summary statistics of the outcomes and covariates} \\
	& \multicolumn{2}{c}{Price increase} & \multicolumn{2}{c}{Tax increase} \\
	\cmidrule(lr){2-3}\cmidrule(lr){4-5}
	& $D=0$ & $D=1$ & $D=0$ & $D=1$ \\ 
	\hline
	\endfirsthead
	
	\caption[]{Summary statistics of the outcomes and covariates (continued)} \\
	& \multicolumn{2}{c}{Price increase} & \multicolumn{2}{c}{Tax increase} \\
	\cmidrule(lr){2-3}\cmidrule(lr){4-5}
	& $D=0$ & $D=1$ & $D=0$ & $D=1$ \\
	\hline
	\endhead
	
	\hline
	\multicolumn{5}{p{\textwidth}}{\footnotesize \textit{(Continued)}} \\ 
	\endfoot
	
	\hline\hline \\
	\caption*{\footnotesize \textit{Note: This table shows the means of outcomes and covariates for the control and treatment groups in the analyses on price and tax increases. Standard deviations are reported in brackets.}}
	\endlastfoot
	
	Smoking rate (current smokers) & 0.26 (0.44) & 0.25 (0.43) & 0.22 (0.42) & 0.23 (0.42) \\ 
	Smoking rate (daily smokers) & 0.25 (0.43) & 0.23 (0.42) & 0.21 (0.41) & 0.22 (0.41) \\ 
	Cigarette prices in PPP\$, real (base $=$ 2020), pre-treatment period & 6.95 (1.97) & 8.51 (2.00) & 8.11 (1.45) & 7.90 (1.80) \\ 
	Taxes as a \% of cigarette price, pre-treatment period & 0.81 (0.04) & 0.77 (0.04) & 0.77 (0.05) & 0.77 (0.04) \\ 
	Age: Age in years & 50.50 (18.38) & 49.87 (17.93) & 51.42 (18.10) & 50.62 (18.26) \\ 
	Age: 99 years and older & 0.00 (0.02) & 0.00 (0.01) & 0.00 (0.01) & 0.00 (0.00) \\ 
	Age (missing) & 0.00 (0.00) & 0.00 (0.01) & 0.00 (0.01) & 0.00 (0.01) \\ 
	Gender: Man & 0.44 (0.50) & 0.47 (0.50) & 0.46 (0.50) & 0.46 (0.50) \\ 
	Gender: Woman & 0.56 (0.50) & 0.53 (0.50) & 0.54 (0.50) & 0.54 (0.50) \\ 
	Gender: Other & 0.00 (0.00) & 0.00 (0.01) & 0.00 (0.00) & 0.00 (0.01) \\ 
	Household composition: 1 person & 0.25 (0.43) & 0.19 (0.40) & 0.24 (0.43) & 0.23 (0.42) \\ 
	Household composition: 2 persons & 0.35 (0.48) & 0.35 (0.48) & 0.37 (0.48) & 0.36 (0.48) \\ 
	Household composition: 3 persons & 0.17 (0.38) & 0.18 (0.39) & 0.17 (0.37) & 0.17 (0.37) \\ 
	Household composition: 4+ persons & 0.23 (0.42) & 0.27 (0.44) & 0.22 (0.42) & 0.24 (0.43) \\ 
	Household composition (missing) & 0.00 (0.03) & 0.00 (0.02) & 0.00 (0.02) & 0.00 (0.02) \\ 
	Foreign nationality & 0.01 (0.09) & 0.04 (0.20) & 0.01 (0.11) & 0.03 (0.17) \\ 
	Age education: Up to 15 & 0.12 (0.33) & 0.14 (0.35) & 0.13 (0.33) & 0.14 (0.35) \\ 
	Age education: 16--19 & 0.39 (0.49) & 0.45 (0.50) & 0.39 (0.49) & 0.42 (0.49) \\ 
	Age education: 20+ & 0.40 (0.49) & 0.33 (0.47) & 0.40 (0.49) & 0.36 (0.48) \\ 
	Age education: Still studying & 0.07 (0.25) & 0.07 (0.25) & 0.06 (0.24) & 0.07 (0.25) \\ 
	Age education: No full-time education & 0.01 (0.12) & 0.00 (0.06) & 0.01 (0.12) & 0.00 (0.07) \\ 
	Age education (missing) & 0.01 (0.10) & 0.01 (0.11) & 0.02 (0.12) & 0.01 (0.11) \\ 
	Marital status: (Re-)Married & 0.50 (0.50) & 0.53 (0.50) & 0.51 (0.50) & 0.49 (0.50) \\ 
	Marital status: Single living with partner & 0.12 (0.33) & 0.10 (0.30) & 0.12 (0.32) & 0.12 (0.32) \\ 
	Marital status: Single & 0.17 (0.37) & 0.18 (0.38) & 0.17 (0.38) & 0.17 (0.38) \\ 
	Marital status: Divorced or separated & 0.09 (0.28) & 0.07 (0.25) & 0.09 (0.29) & 0.08 (0.27) \\ 
	Marital status: Widowed & 0.10 (0.31) & 0.08 (0.28) & 0.10 (0.30) & 0.10 (0.30) \\ 
	Marital status: Other & 0.01 (0.11) & 0.01 (0.08) & 0.01 (0.08) & 0.01 (0.08) \\ 
	Marital status (missing) & 0.00 (0.05) & 0.03 (0.17) & 0.00 (0.05) & 0.03 (0.17) \\ 
	Type of community: Rural area or village & 0.29 (0.45) & 0.34 (0.47) & 0.32 (0.47) & 0.36 (0.48) \\ 
	Type of community: Small/middle town & 0.41 (0.49) & 0.34 (0.47) & 0.39 (0.49) & 0.35 (0.48) \\ 
	Type of community: Large town & 0.30 (0.46) & 0.32 (0.47) & 0.30 (0.46) & 0.29 (0.45) \\ 
	Type of community (missing) & 0.00 (0.02) & 0.00 (0.02) & 0.00 (0.03) & 0.00 (0.02) \\ 
	Occupation: Self-employed & 0.07 (0.26) & 0.08 (0.27) & 0.07 (0.25) & 0.07 (0.26) \\ 
	Occupation: Managers & 0.09 (0.29) & 0.10 (0.31) & 0.11 (0.32) & 0.11 (0.32) \\ 
	Occupation: Other white collars & 0.11 (0.31) & 0.13 (0.34) & 0.13 (0.34) & 0.12 (0.33) \\ 
	Occupation: Manual workers & 0.19 (0.39) & 0.20 (0.40) & 0.20 (0.40) & 0.19 (0.39) \\ 
	Occupation: House persons & 0.04 (0.20) & 0.07 (0.25) & 0.04 (0.19) & 0.05 (0.22) \\ 
	Occupation: Unemployed & 0.08 (0.28) & 0.06 (0.24) & 0.06 (0.24) & 0.06 (0.24) \\ 
	Occupation: Retired & 0.34 (0.47) & 0.29 (0.45) & 0.33 (0.47) & 0.32 (0.46) \\ 
	Occupation: Students & 0.07 (0.25) & 0.07 (0.25) & 0.06 (0.24) & 0.07 (0.25) \\ 
	MPOWER indicator P level 2 & 0.25 (0.44) & 0.33 (0.47) & 0.56 (0.50) & 0.39 (0.49) \\ 
	MPOWER indicator P level 3 & 0.50 (0.50) & 0.22 (0.41) & 0.11 (0.31) & 0.33 (0.47) \\ 
	MPOWER indicator P level 4 & 0.13 (0.33) & 0.08 (0.27) & 0.11 (0.32) & 0.08 (0.27) \\ 
	MPOWER indicator P level 5 & 0.12 (0.33) & 0.37 (0.48) & 0.22 (0.41) & 0.20 (0.40) \\
	MPOWER indicator O level 4 & 0.88 (0.33) & 0.74 (0.44) & 0.61 (0.49) & 0.89 (0.32) \\ 
	MPOWER indicator O level 5 & 0.12 (0.33) & 0.26 (0.44) & 0.39 (0.49) & 0.11 (0.32) \\ 
	MPOWER indicator W level 3 & 0.37 (0.48) & 0.19 (0.39) & 0.22 (0.41) & 0.30 (0.46) \\ 
	MPOWER indicator W level 4 & 0.25 (0.43) & 0.12 (0.33) & 0.22 (0.41) & 0.11 (0.31) \\ 
	MPOWER indicator W level 5 & 0.38 (0.48) & 0.69 (0.46) & 0.56 (0.50) & 0.58 (0.49) \\ 
	MPOWER indicator E level 4 & 0.87 (0.34) & 0.89 (0.31) & 1.00 (0.00) & 0.89 (0.31) \\ 
	MPOWER indicator E level 5 & 0.13 (0.34) & 0.11 (0.31) & 0.00 (0.00) & 0.11 (0.31) 
\end{longtable}

\newpage

% table main results

\begin{table}[htbp!]
	\caption{Effect of price and tax increases on the smoking rate} 
	\label{tab:mainRes}
	\centering
	\scriptsize
	\begin{tabular}{p{.12\textwidth}p{.08\textwidth}p{.08\textwidth}p{.08\textwidth}p{.08\textwidth}p{.08\textwidth}p{.08\textwidth}p{.08\textwidth}p{.08\textwidth}}
		& \multicolumn{5}{c}{DiDDML with random forests} & \multicolumn{3}{c}{Parametric DiD} \\
		\cmidrule(lr){2-6}\cmidrule(lr){7-9}
		& (1) & (2) & (3) & (4) & (5) & (6) & (7) & (8) \\ 
		\hline
		\multicolumn{9}{l}{Effect of price increases} \\ 
		\hline
		\multicolumn{9}{l}{Current smokers} \\ 
		ATET & -0.0073 & -0.0037 & 0.0038 & 0.0017 & 0.0017 & 0.0032 & -0.0066 & -0.0029 \\  
		Std. error & (0.0113) & (0.0093) & (0.0107) & (0.0080) & (0.0116) & (0.0105) & (0.0030) & (0.0026) \\ 
		P-value & 0.5166 & 0.6929 & 0.7191 & 0.8265 & 0.8799 & 0.7570 & 0.0263 & 0.2748 \\ 
		N & 53,891 & 53,891 & 53,891 & 53,891 & 53,891 & 53,891 & 53,891 & 107,258 \\ 
		Trimmed & 0 & 0 & 0 & 0 & 0 & 0 & 0 & 0\\ 
		\multicolumn{9}{l}{Daily smokers} \\ 
		ATET & -0.0004 & 0.0031 & 0.0045 & 0.0056 & 0.0056 & 0.0047 & -0.0060 & -0.0038 \\ 
		Std. error & (0.0107) & (0.0087) & (0.0108) & (0.0082) & (0.0113) & (0.0095) & (0.0035) & (0.0029) \\ 
		P-value & 0.9724 & 0.7254 & 0.6763 & 0.4910 & 0.6179 & 0.6256 & 0.0917 & 0.1895 \\
		N & 53,891 & 53,891 & 53,891 & 53,891 & 53,891 & 53,891 & 53,891 & 107,258 \\
		Trimmed & 0 & 0 & 0 & 0 & 0 & 0 & 0 & 0 \\
		\hline
		\multicolumn{9}{l}{Effect of tax increases} \\ 
		\hline
		\multicolumn{9}{l}{Current smokers} \\ 
		ATET & -0.0136 & -0.0245 & -0.0214 & -0.0344 & -0.0344 & -0.0225 & -0.3977 & -0.1163 \\
		Std. error & (0.0075) & (0.0087) & (0.0130) & (0.0167) & (0.0137) & (0.0072) & (0.1297) & (0.0635) \\   
		P-value & 0.0698 & 0.0050 & 0.1002 & 0.0392 & 0.0118 & 0.0017 & 0.0022 & 0.0671 \\ 
		N & 55,329 & 55,329 & 55,329 & 55,329 & 55,329 & 55,329 & 55,329 & 107,258 \\ 
		Trimmed & 0 & 0 & 0 & 0 & 0 & 0 & 0 & 0 \\ 
		\multicolumn{9}{l}{Daily smokers}  \\ 
		ATET & -0.0162 & -0.0272 & -0.0222 & -0.0315 & -0.0315 & -0.0232 & -0.3568 & -0.1127 \\  
		Std. error & (0.0080) & (0.0102) & (0.0120) & (0.0184) & (0.0128) & (0.0072) & (0.1287) & (0.0734) \\  
		P-value & 0.0435 & 0.0074 & 0.0653 & 0.0870 & 0.0140 & 0.0014 & 0.0056 & 0.1245 \\ 
		N & 55,329 & 55,329 & 55,329 & 55,329 & 55,329 & 55,329 & 55,329 & 107,258 \\
		Trimmed & 0 & 0 & 0 & 0 & 0 & 0 & 0 & 0 \\
		\hline
		Treat. hist. & X & X & X & X & X & X & X & X\\ 
		Socio-dem. &  & X &  & X & X & X & X & X\\ 
		Policies &  &  & X & X & X & X & X & X\\ 
		Clustering & X & X & X & X &  & X & X & X\\ 
		\hline\hline
	\end{tabular}
	\caption*{\footnotesize \textit{Note: The first panel shows the effect of price increases on the smoking rate, comparing the treatment group (15\%+ increase) with the control group ($\pm$5\% change). The second panel shows the effect of tax increases on the smoking rate, comparing the treatment group (2\%+ increase) with the control group (no change). Column (4) presents the main results estimated using DiDDML with random forests, including all covariates and cluster-robust standard errors. The remaining columns report additional analyses that vary the covariate composition, clustering, and method.}}
\end{table}

\newpage

% table heterogeneous results

\begin{table}[htbp!]
	\caption{Effect of price and tax increases on the smoking rate, by socio-demographic groups}
	\label{tab:hetRes}
	\centering
	\scriptsize
	\begin{tabular}{p{.12\textwidth}p{.07\textwidth}p{.07\textwidth}p{.07\textwidth}p{.07\textwidth}p{.07\textwidth}p{.07\textwidth}p{.07\textwidth}p{.07\textwidth}p{.07\textwidth}}
		& \multicolumn{2}{c}{Gender} & \multicolumn{4}{c}{Age} & \multicolumn{3}{c}{Age last education} \\
		\cmidrule(lr){2-3} \cmidrule(lr){4-7} \cmidrule(lr){8-10}
		& Men & Women & 15--24 & 25--44 & 45--64 & 65+ & $\leq$15 & 16--19 & 20+ \\ 
		\hline
		\multicolumn{10}{l}{Effect of price increases}  \\ 
		\hline
		\multicolumn{10}{l}{Current smokers}  \\
		ATET & -0.0084 & 0.0106 & 0.0540 & -0.0074 & 0.0235 & -0.0063 & 0.0487 & 0.0143 & -0.0083 \\ 
		Std. error & (0.0189) & (0.0136) & (0.0259) & (0.0166) & (0.0204) & (0.0140) & (0.0266) & (0.0191) & (0.0096) \\ 
		P-value & 0.6553 & 0.6553 & 0.3024 & 0.6553 & 0.6553 & 0.6553 & 0.3024 & 0.6553 & 0.6553 \\ 
		N & 24,837 & 29,051 & 5,321 & 15,898 & 18,917 & 13,743 & 7,418 & 23,122 & 18,736 \\ 
		Trimmed & 1 & 0 & 0 & 0 & 0 & 0 & 0 & 0 & 0 \\
		\multicolumn{10}{l}{Daily smokers}  \\ 
		ATET & -0.0016 & 0.0083 & 0.0522 & -0.0005 & 0.0226 & -0.0074 & 0.0408 & 0.0132 & 0.0000 \\ 
		Std. error & (0.0178) & (0.0139) & (0.0242) & (0.0179) & (0.0202) & (0.0120) & (0.0270) & (0.0209) & (0.0073) \\ 
		P-value & 0.9995 & 0.8272 & 0.2763 & 0.9995 & 0.7902 & 0.8272 & 0.5908 & 0.8272 & 0.9995 \\ 
		N & 24,837 & 29,051 & 5,321 & 15,898 & 18,917 & 13,743 & 7,418 & 23,122 & 18,736 \\ 
		Trimmed & 1 & 0 & 0 & 0 & 0 & 0 & 0 & 0 & 0 \\
		\hline
		\multicolumn{10}{l}{Effect of tax increases}  \\ 
		\hline
		\multicolumn{10}{l}{Current smokers}  \\
		ATET & -0.0246 & -0.0426 & -0.1066 & -0.0021 & -0.0609 & -0.0055 & -0.0614 & -0.0456 & -0.0170 \\ 
		Std. error & (0.0315) & (0.0175) & (0.0333) & (0.0223) & (0.0294) & (0.0085) & (0.0289) & (0.0225) & (0.0131) \\ 
		P-value & 0.5592 & 0.0684 & 0.0126 & 0.9251 & 0.0763 & 0.5879 & 0.0763 & 0.0763 & 0.2914 \\ 
		N & 25,366 & 29,961 & 5,181 & 15,622 & 19,572 & 14,950 & 7,498 & 22,651 & 20,424 \\ 
		Trimmed & 0 & 0 & 0 & 0 & 0 & 0 & 0 & 0 & 0 \\
		\multicolumn{10}{l}{Daily smokers}  \\
		ATET & -0.0284 & -0.0378 & -0.1026 & -0.0039 & -0.0462 & -0.0067 & -0.0610 & -0.0425 & -0.0172 \\ 
		Std. error & (0.0295) & (0.0195) & (0.0353) & (0.0225) & (0.0331) & (0.0090) & (0.0323) & (0.0229) & (0.0128) \\ 
		P-value & 0.4312 & 0.1424 & 0.0333 & 0.8642 & 0.2692 & 0.5105 & 0.1424 & 0.1424 & 0.2692 \\ 
		N & 25,366 & 29,961 & 5,181 & 15,622 & 19,572 & 14,950 & 7,498 & 22,651 & 20,424 \\ 
		Trimmed & 0 & 0 & 0 & 0 & 0 & 0 & 0 & 0 & 0 \\
		\hline\hline
	\end{tabular}
	\caption*{\footnotesize \textit{Note: The first panel shows the effect of price increases on the smoking rate, comparing the treatment group (15\%+ increase) with the control group ($\pm$5\% change), by socio-demographic groups. The second panel shows the effect of tax increases on the smoking rate, comparing the treatment group (2\%+ increase) with the control group (no change), by socio-demographic groups. All results are based on the main analysis specification, i.e., DiDDML with random forests, including all covariates and cluster-robust standard errors. P-values are adjusted for multiple hypothesis testing using the Benjamini--Hochberg method \parencite{Benjamini95}.}}
\end{table}

\newpage

% table results in 2012-2014 and 2018-2020

\begin{table}[htbp!]
	\caption{Effect of price and tax increases on the smoking rate, in 2012--2014 and 2018--2020}
	\label{tab:mainRes12141820}
	\centering
	\scriptsize
	\begin{tabular}{p{.12\textwidth}p{.12\textwidth}p{.12\textwidth}p{.12\textwidth}p{.12\textwidth}}
		& \multicolumn{2}{c}{Price increase} & \multicolumn{2}{c}{Tax increase} \\
		\cmidrule(lr){2-3} \cmidrule(lr){4-5}
		& 2012--2014 & 2018--2020 & 2012--2014 & 2018--2020 \\
		\hline
		\multicolumn{5}{l}{Current smokers} \\
		\hline
		ATET & -0.0120 & 0.0120 & -0.0138 & -0.0364 \\ 
		Std. error & (0.0121) & (0.0117) & (0.0120) & (0.0191) \\ 
		P-value & 0.3209 & 0.3062 & 0.2490 & 0.0573 \\ 
		N & 21,973 & 31,918 & 23,390 & 31,939 \\ 
		Trimmed & 0 & 0 & 0 & 0 \\ 
		\multicolumn{5}{l}{Daily smokers} \\
		\hline
		ATET & -0.0188 & 0.0206 & -0.0148 & -0.0327 \\ 
		Std. error & (0.0137) & (0.0125) & (0.0133) & (0.0223) \\ 
		P-value & 0.1717 & 0.0994 & 0.2673 & 0.1421 \\ 
		N & 21,973 & 31,918 & 23,390 & 31,939 \\ 
		Trimmed & 0 & 0 & 0 & 0 \\ 
		\hline\hline
	\end{tabular}
	\caption*{\footnotesize \textit{Note: This table shows the effect of price and tax increases on the smoking rate, estimating the main model only based on observations from period 2012--2014 or 2018--2020. All results are based on the main analysis specification, i.e., DiDDML with random forests, including all covariates and cluster-robust standard errors.}}
\end{table}

% plot treatment propensitiy scores

\begin{figure}[htbp!]
	\centering
	\subfigure[Price increase]{\includegraphics[width=0.45\textwidth]{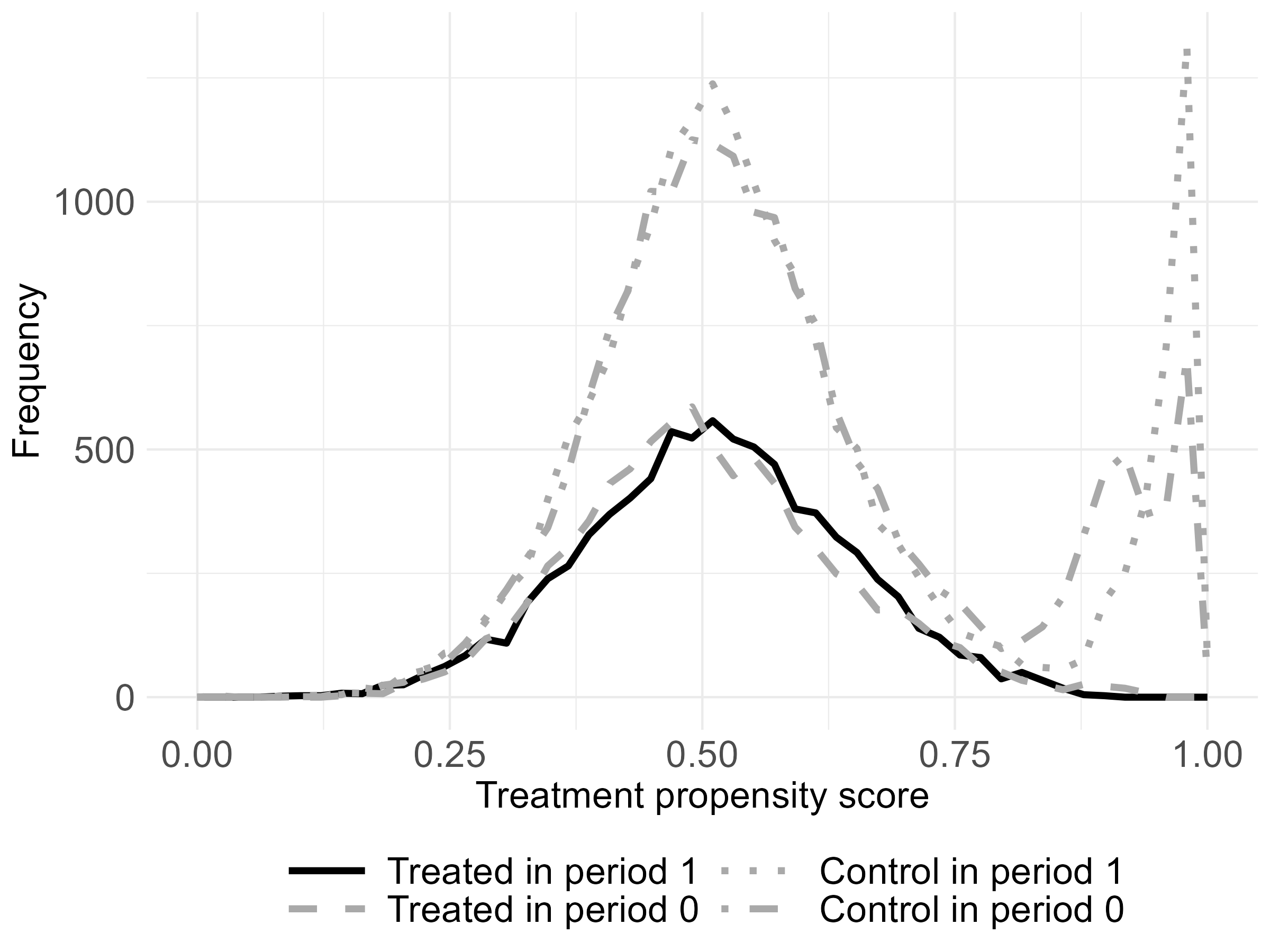}}
	\subfigure[Tax increase]{\includegraphics[width=0.45\textwidth]{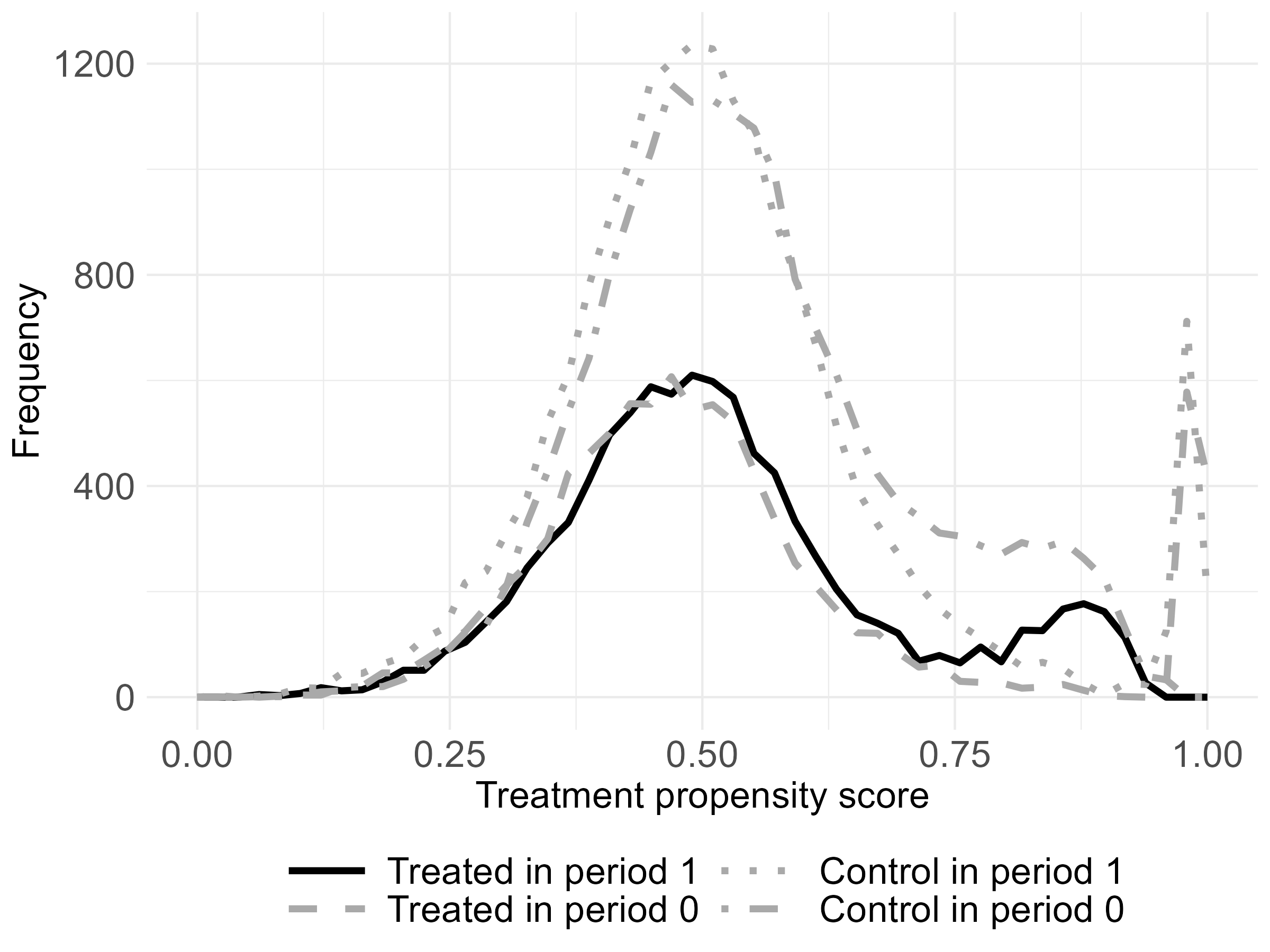}}
	\caption{Treatment propensity scores}
	\label{fig:propensityScores}
	\caption*{\footnotesize \textit{Note: The treatment propensity scores are based on the main analysis specification using DiDDML with random forests, including all covariates. The distribution of propensity scores is shown for all four comparison groups: (i) under treatment in period 1 (solid line), (ii) under treatment in period 0 (dashed line), (iii) under control in period 1 (dotted line), and (iv) under control in period 0 (dash-dotted line).}}
\end{figure}

\newpage

% placebo effects

\begin{figure}[htbp!]
	\centering
	\subfigure[Price increase -- current smokers]{\includegraphics[width=0.45\textwidth]{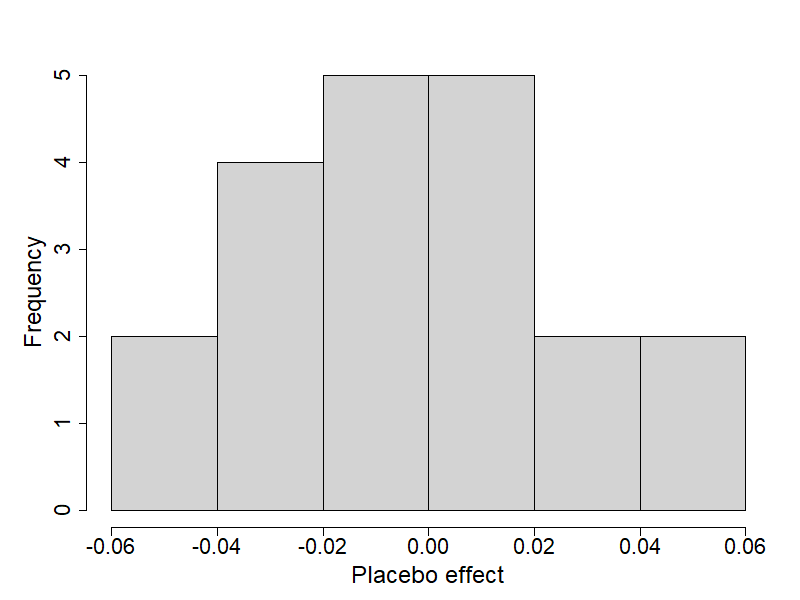}}
	\subfigure[Price increase -- daily smokers]{\includegraphics[width=0.45\textwidth]{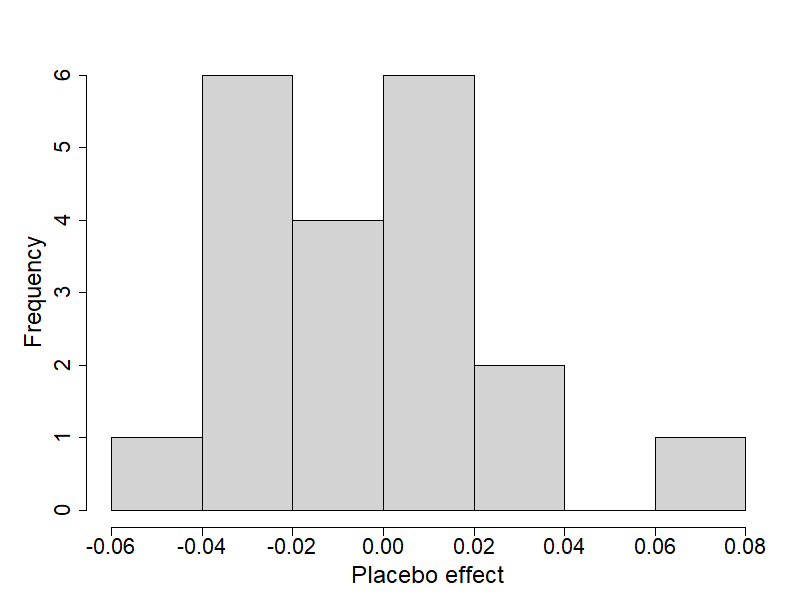}}
	\subfigure[Tax increase -- current smokers]{\includegraphics[width=0.45\textwidth]{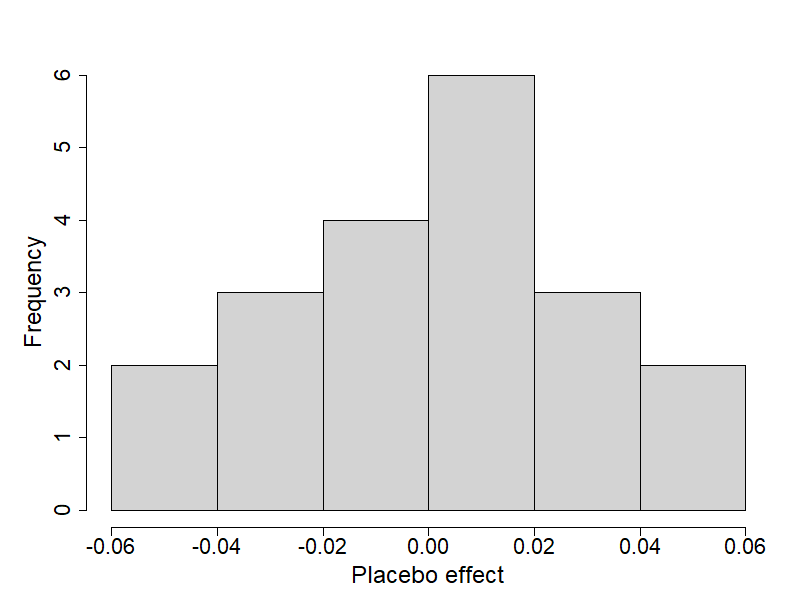}}
	\subfigure[Tax increase -- daily smokers]{\includegraphics[width=0.45\textwidth]{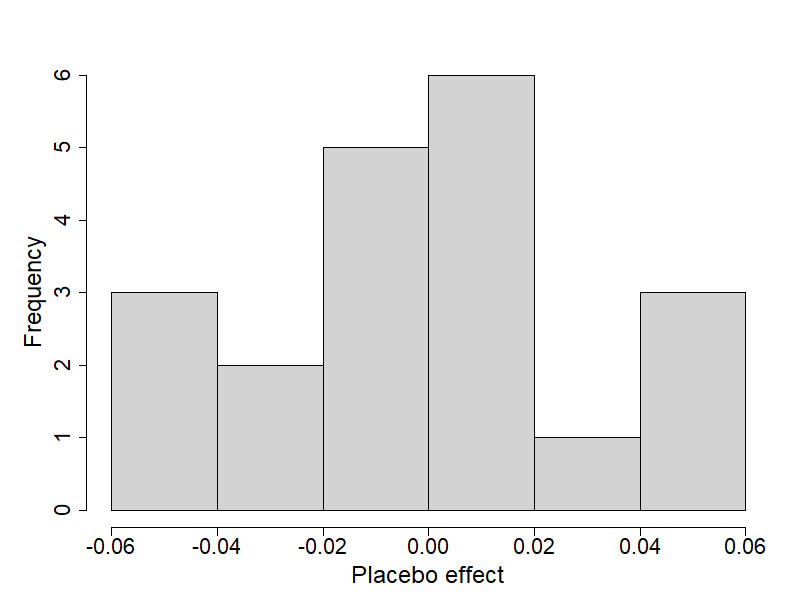}}
	\caption{Distribution of placebo effects}
	\label{fig:placeboTests}
	\caption*{\footnotesize \textit{Note: The plots present the distributions of the placebo effects for each of the four main analyses as described in \cref{sec:robustChecks}.}}
\end{figure}
	
\end{appendices}

\end{document}